\documentstyle[epsf,floats,tighten,preprint,aps]{revtex}
\preprint{RIKEN-AF-NP-218}
\renewcommand{\vec}[1]{{\bf #1}}
\newcommand{\calcium}{{}^{40}{\rm Ca}}

\title{
Antisymmetrized molecular dynamics of wave packets\\
with stochastic incorporation of Vlasov equation }
\author{Akira Ono}
\address{Institute of Physical and Chemical Research (RIKEN), Wako,
Saitama 351-01, Japan}
\author{Hisashi Horiuchi}
\address{Department of Physics, Kyoto University, Kyoto 606-01, Japan}
\draft
\begin{document}
\maketitle
\begin{abstract}
On the basis of the antisymmetrized molecular dynamics (AMD) of wave
packets for the quantum system, a novel model (called AMD-V) is
constructed by the stochastic incorporation of the diffusion and the
deformation of wave packets which is calculated by Vlasov equation
without any restriction on the one-body distribution. In other words,
the stochastic branching process in molecular dynamics is formulated
so that the instantaneous time evolution of the averaged one-body
distribution is essentially equivalent to the solution of Vlasov
equation. Furthermore, as usual molecular dynamics, AMD-V keeps the
many-body correlation and can naturally describe the fluctuation among
many channels of the reaction. It is demonstrated that the newly
introduced process of AMD-V has drastic effects in heavy ion
collisions of $\calcium+\calcium$ at 35 MeV/nucleon, especially on the
fragmentation mechanism, and AMD-V reproduces the fragmentation data
very well. Discussions are given on the interrelation among the
frameworks of AMD, AMD-V and other microscopic models developed for
the nuclear dynamics.
\end{abstract}

\narrowtext
\section{Introduction}

Heavy ion reactions in medium energy region give us opportunities to
study the dynamics of the highly excited nuclear system far from the
equilibrium and various kinds of theoretical models have been
developed to describe the variety of phenomena realized in such
reactions. There are many models which belong to the category of
one-body transport models, such as the time-dependent Hartree Fock
theory (TDHF). Vlasov equation {}\cite{WONG} can be regarded as the
semiclassical approximation of TDHF. In these models, the one-body
distribution function for a Slater determinant propagates in the mean
field which depends on itself. Vlasov-Uehling-Uhlenbeck (VUU) equation
{}\cite{BERTSCH,CASSING} includes the two-nucleon collision term as
the effect of the residual interaction on the one-body distribution
function. However, the description only with a one-body distribution
function cannot be a good framework for the reaction system which has
various final channels like fragmentation and therefore should be
described by the linear combination of many Slater determinants in
principle. When the one-body transport models are applied to such a
system, the use of the averaged mean field common to all channels
brings about spurious correlation among the time evolutions of
channels which should be independent of one another in the true
solution due to the linearity of the time-dependent Schr\"odinger
equation.

On the other hand, molecular dynamics models
{}\cite{AICHELIN,BOAL,MARUb,FELDMEIER,ONOab} can be applied to the
reaction system with many channels because they treat each channel as
an independent event. Although the one-body transport model is an
approximation of the molecular dynamics for point particles in the
context of the classical dynamics, this relation is rather opposite
for the quantum system in which any particle should have phase space
distribution due to the uncertainty relation. In the antisymmetrized
molecular dynamics (AMD) {}\cite{ONOab}, for example, the system is
described by a Slater determinant of Gaussian wave packets and the
time evolution of the centroids of wave packets is determined by the
equation of motion derived from the time-dependent variational
principle. This equation of motion may be interpreted as an
approximation of TDHF (or Vlasov) equation in the sense that TDHF
gives better prediction of the instantaneous time evolution of an AMD
wave function than AMD does because the TDHF single-particle wave
function is more general than the AMD single-particle wave
function. However, it should be noticed that this does not always mean
that TDHF is superior to AMD in the global time scale, because AMD
treats many channels and respects their independence, which is more
important than the flexibility of single-particle wave functions for
the system with many channels such as the multifragmentation. In AMD
and many other molecular dynamics models, the branching process into
channels is introduced by the stochastic collisions between two wave
packets which correspond to the collision term in VUU equation but
produce large fluctuation among events or channels. Therefore, even
when we begin with a single initial state, many final channels are
obtained by repeating the calculation. It is generally believed that
molecular dynamics models can describe the fragmentation in heavy ion
collisions.

Besides the two-nucleon collision process, there is another source of
branching which is important even in low energy phenomena irrelevant
to the two-nucleon collisions. A simple example can be found in the
nucleon emission from a hot nucleus. When TDHF is applied to a hot
fragment which has been produced in one of the channels of the true
solution, a minor part of a single particle wave function will go out
of the nucleus and the other main part will remain in the nucleus
unless the excitation energy is very large. Although TDHF will give
good predictions of the emission probability and the nucleon spectrum,
how reliable is the description of the residue nucleus? Depending on
whether the nucleon has been emitted or not, the mean field of the
residue nucleus should be made by $A-1$ nucleons or $A$ nucleons
respectively, but TDHF and other one-body transport models do not
include such fluctuation between channels. Similar kind of fluctuation
among channels is also important in the nucleon transfer in heavy ion
collision, the fragment formation by the coalescence mechanism, and so
on. In many cases when a single-particle wave function has spread wide
or has splitted into several parts, the system should be branching
into channels and therefore the TDHF description as a Slater
determinant breaks down. In such cases, it seems better to decompose a
Slater determinant into several Slater determinants and solve the
later time evolution of each component independently.

On the other hand, in the molecular dynamics models with wave packets,
the equation of motion chooses only the most important channel among
the channels of the true solution if there is no source of fluctuation
such as the stochastic two-nucleons collisions, though a variety of
final channels are possibly generated in medium energy heavy ion
collisions due to the fluctuation produced by the two-nucleon
collisions. In our recent study of statistical property of AMD
{}\cite{ONOf,ONOg}, we found that the ensemble of AMD wave functions
of a hot nucleus has good statistical property of quantum mechanics in
the observables such as the single-particle momentum distribution and
the occupation probability of single-particle levels
{}\cite{ONOg}. However, AMD has problem in the description of the
future time evolution of the minor high-momentum component as an
independent branch, which is the origin of the failure in the nucleon
emission and the phase equilibrium of liquid and gas
{}\cite{ONOf}. The minor branch of the nucleon emission which should
be caused by the high-momentum tail is not respected at all in the
dynamics and the whole nucleon wave packet remains in the nucleus
because the wave packet is not allowed to split off.  Therefore the
ironical conclusion of the study of statistical property of AMD was
that the problem is not due to anything very complex and
uncontrollable by the usual microscopic considerations but due to the
rather simple single-particle motion which has restriction in
molecular dynamics with wave packets. This means in turn that we have
chance to overcome this problem by respecting the spreading and the
splitting of wave packets which are naturally predicted by simple
one-body considerations. In fact, we showed in Ref.\ {}\cite{ONOf}
that the problem of the nucleon emission can be solved by taking
account of the stochastic splitting of the wave packet based on its
momentum width when a nucleon wave packet is being emitted from the
nuclear surface. It should be emphasized that with this stochastic
method there is no spurious correlation of the above-mentioned TDHF
calculation between channels with and without nucleon
emission. However, in the dynamics of nuclear reactions, there may be
other phenomena caused by the wave packet tail that are lost
completely in AMD due to the restriction of the single-particle
states.

The first purpose of this paper is to present an extended AMD model
which can generally describe such minor branching processes by
removing the restriction on the one-body distribution function.  This
is done not by generalizing the wave packets to arbitrary
single-particle wave functions but by representing the diffused and/or
deformed wave packet as an ensemble of Gaussian wave packets. In other
words, stochastic displacements are given to the wave packets in phase
space so that the ensemble-average of the time evolution of the
one-body distribution function is essentially equivalent to the
solution of Vlasov equation which does not have any restriction on the
shape of wave packets. This new model is called AMD-V. Although AMD-V
is equivalent to Vlasov equation in the instantaneous time evolution
of the one-body distribution function for an AMD wave function, AMD-V
describes the branching into channels and the fluctuation of the mean
field which are caused by the spreading or the splitting of the
single-particle wave function.  Furthermore, the stochastic
two-nucleon collision process is included in AMD-V just in the same
way as in AMD and other molecular dynamics models.

The second purpose of this paper is to show the drastic effect of this
new stochastic process of wave packet splitting on the dynamics of
heavy ion collisions, especially in the fragmentation mechanism. We
take the $\calcium+\calcium$ system at the incident energy 35
MeV/nucleon. It will be shown that the reproduction of data by the
AMD-V calculation is surprisingly good. From the previous study of the
nucleon-emission process, it is automatically expected that the decay
of the produced fragments should be too slow in AMD calculation, which
will be proved in this paper by the comparison of AMD and
AMD-V. However, the deviation between AMD and AMD-V appears not only
in the decay of equilibrated fragments but also in early stages of the
reaction. In fact, we will see that the effect of the wave packet
diffusion is crucially important to remove the spurious binary feature
of the AMD calculation and to enable the multi-fragment final state.

This paper is organized as follows. In Sec.\ II, the formulation of
the new process of AMD-V is given in detail, and then in Sec.\ III,
the calculated results of AMD and AMD-V are compared with each other
and also with the experimental fragmentation data, to show the
importance of the new process of AMD-V. Section IV is devoted to the
discussion on the relation of AMD and AMD-V to other microscopic
dynamical models such as TDHF and VUU. Summary is given in Sec.\ V.

\section{Formulation of AMD-V}
\subsection{Usual AMD}

Before the incorporation of the stochastic process of wave packet
splitting, we will explain the usual AMD {}\cite{ONOab} very briefly
for the convenience of the readers. AMD describes the nuclear many
body system by a Slater determinant of Gaussian wave packets as
\begin{equation}
\Phi(Z)=
\det\Bigl[\exp\Bigl\{-\nu({\vec r}_j - {\vec Z}_i/\sqrt
\nu)^2+{1\over2}{\vec Z}_i^2\Bigr\}
\chi_{\alpha_i}(j)\Bigr], 
\end{equation}
where the complex variables $Z\equiv\{{\vec Z}_i\}$ are the centroids
of the wave packets. We took the width parameter $\nu=0.16\,{\rm
fm}^{-2}$ and the spin isospin states $\chi_{\alpha_i}={\rm
p}\uparrow$, ${\rm p}\downarrow$, ${\rm n}\uparrow$, or ${\rm
n}\downarrow$.  The time evolution of $Z$ is determined by the
time-dependent variational principle and the two-nucleon collision
process. The equation of motion for $Z$ derived from the
time-dependent variational principle is
\begin{equation}
  i\hbar\sum_{j\tau}C_{i\sigma,j\tau}{dZ_{j\tau}\over dt}=
  {\partial{\cal H}\over\partial Z_{i\sigma}^*}.
  \label{eq:AMDEqOfMotion}
\end{equation}
$C_{i\sigma,j\tau}$ with $\sigma,\tau=x,y,z$ is a hermitian matrix,
and $\cal H$ is the expectation value of the Hamiltonian after the
subtraction of the spurious kinetic energy of the zero-point
oscillation of the center-of-masses of fragments,
\begin{equation}
  {\cal H}(Z)={\langle\Phi(Z)|H|\Phi(Z)\rangle
               \over\langle\Phi(Z)|\Phi(Z)\rangle}
          -{3\hbar^2\nu\over2M}A+T_0(A-N_{\rm F}(Z)),
  \label{eq:AMDHamil}
\end{equation}
where $N_{\rm F}(Z)$ is the fragment number, and $T_0$ is
$3\hbar^2\nu/2M$ in principle but treated as a free parameter for the
adjustment of the binding energies. Two-nucleon collisions are
introduced by the use of the physical coordinates $W=\{{\vec W}_i\}$
which are defined as
\begin{equation}
  {\vec W}_i=\sum_{j=1}^A \Bigl(\sqrt Q\Bigr)_{ij}{\vec Z}_j,\quad
  Q_{ij} ={\partial \log \langle\Phi(Z)|\Phi(Z)\rangle
           \over \partial({\vec Z}_i^*\cdot{\vec Z}_j)}.
  \label{eq:PhysCoord}
\end{equation}

\subsection{Stochastic incorporation of Vlasov equation}

\subsubsection{Basic idea}

In molecular dynamics models with wave packets, each nucleon $i$ at
the time $t=t_0$ is represented by a Gaussian wave packet in phase
space
\begin{equation}
f_i({\vec r},{\vec p},t_0)=8\, e^{
  -2\nu({\vec r}-{\vec R}_i(t_0))^2
  -({\vec p}-{\vec P}_i(t_0))^2/2\hbar^2\nu
  },
\end{equation}
with the centroid ${\vec R}_i$ and ${\vec P}_i$. This wave packet
satisfies the minimum uncertainty relation. The total one-body
distribution function $f$ is the sum of $f_i$. In the case of AMD,
this representation of each nucleon as a simple Gaussian wave packet
is valid approximately if we use the physical coordinate
\begin{equation}
{\vec W}_i=\sqrt\nu\,{\vec R}_i+{i\over2\hbar\sqrt\nu}{\vec P}_i
\end{equation}
as the centroid. The time evolution of the centroids $\dot{\vec R}_i$
and $\dot{\vec P}_i$ are derived from the equation of motion while the
shape of wave packets is fixed.

However, more reliable time evolution of the one-body distribution
function is given by TDHF equation or Vlasov equation, which is
the semiclassical approximation of TDHF equation,
\begin{equation}
{\partial f_i\over\partial t}
+{\partial h\over\partial{\vec p}}
  \cdot{\partial f_i\over\partial{\vec r}}
-{\partial h\over\partial{\vec r}}
  \cdot{\partial f_i\over\partial{\vec p}}
=0,
\end{equation}
where $h=h({\vec r},{\vec p},t)$ is the Wigner representation of the
single-particle Hamiltonian calculated for the AMD wave function
$\Phi(Z(t_0))$ which is a Slater determinant. Although Vlasov (or
TDHF) equation cannot give the reliable time evolution of the one-body
distribution function in the situation where the system has branched
into many channels like the fragmentation in heavy ion collisions,
what we assume here is that the system is represented at $t=t_0$ by an
AMD wave function $\Phi(Z(t_0))$ which is a Slater determinant of
compact single-particle wave functions, and therefore we can safely
trust Vlasov equation for the instantaneous time evolution of the
one-body distribution function.

In order to reflect Vlasov equation to AMD, we take the following
stochastic procedure for each nucleon $i$ during the short time step
between $t_0$ and $t_0+\delta t$. For the simplicity of formulae, we
introduce new notations
\begin{eqnarray}
&& x=\{x_a\}_{a=1,\ldots,6}
    =\Bigl\{\sqrt\nu\,{\vec r},\; {\vec p}/2\hbar\sqrt\nu\Bigr\},\\
&& X_i=\{X_{ia}\}_{a=1,\ldots,6}
    =\{{\vec W}_i\}
    =\Bigl\{\sqrt\nu\,{\vec R}_i,\; {\vec P}_i/2\hbar\sqrt\nu\Bigr\}.
\end{eqnarray}
Then the one-body distribution function at $t=t_0$ is represented as
\begin{eqnarray}
&& f_i(x,t_0)=F(x-X_i(t_0)),\\
&& F(x)=\prod_{a=1}^6\sqrt{2/\pi}\,e^{-2x_a^2}.
\end{eqnarray}
The essential point of AMD-V is to write the one-body distribution
function at $t=t_0+\delta t$ as a superposition of Gaussian functions
as
\widetext
\begin{equation}
f_i(x,t_0+\delta t)= (1-c)F\Bigl(x-X_i(t_0+\delta t)\Bigr)
            +c\int g(\xi)F\Bigl(x-X_i(t_0+\delta t)-\xi\Bigr)d\xi ,
\label{eq:f-diffuse}
\end{equation}
\narrowtext\noindent
with the integration variables $\xi=\{\xi_a\}_{a=1,\ldots,6}$. Here we
have introduced a parameter $c$ and a normalized function $g(\xi)$
which depend on $\Phi(Z(t_0))$, $\delta t$ and $i$. The case of $c=0$
corresponds to the usual AMD without shape changes of wave packets. If
we allow arbitrary $g(\xi)$ and $c$, it will be always possible to
represent the exact solution of Vlasov equation. In order to enable
the following prescription, it is further necessary to assume
$g(\xi)\ge 0$ and $0\le c\le 1$. This restriction disables the
description of the shrinking of the wave packet but seems reasonable
since even under this restriction it is possible to describe the
diffusion of the wave packet which is the important origin of the
branching into channels but missing in the usual AMD. Let us assume
that $g(\xi)$ and $c$ have been determined with a method given later
so as to reproduce the solution of Vlasov equation as much as
possible.  Then it is possible to reflect Eq.\ (\ref{eq:f-diffuse})
exactly within the framework of AMD by giving the stochastic
displacement $\xi$ with the probability $c$ to the centroid of the
wave packet according to the distribution function $g(\xi)$, together
with the usual time evolution of the centroid by the equation of
motion (and the stochastic two-nucleon collisions).  It should be
emphasized that the average value of the one-body distribution
function after this stochastic process is just the same as Eq.\
(\ref{eq:f-diffuse}) and the stochastic implementation is not an
approximate treatment except for the restriction of $g(\xi)\ge0$ and
$0\le c\le 1$.

For the time step between $t_0$ and $t_0+\delta t$, the
above-mentioned stochastic procedures are taken for all nucleons
$i$. This means that the system has changed into an ensemble of many
branches (or channels) at $t_0+\delta t$, while it was a single Slater
determinant at $t_0$. Each branch is represented by an AMD wave
function and will make further branching in the following time steps
just in the same way as was done at $t_0$. The future time evolution
of each branch is solved without any influence from other
branches. What is decisively important here is that the mean fields in
$h$ are different from branch to branch, and therefore the fluctuation
among channels are treated correctly unlike TDHF and other one-body
transport models. This situation is just the same as the branching
caused by the stochastic two-nucleon collisions. In the practical
calculation, only a path of branchings is followed in each simulation,
and many simulations are repeated to get any observable which is
calculated as the ensemble-average value of the expectation value all
over the simulations.

\subsubsection{Practical determination of the stochastic displacement}

Now we explain a method to determine $g(\xi)$ and $c$ which we take in
the calculation to be presented in this paper. Although there can be
various methods, we take here the simplest method by taking account of
only the dispersion of the wave packet
\begin{eqnarray}
&&
\sigma_{ab}^2(t)=\int \Bigl(x_a-\bar X_a(t)\Bigr)
                       \Bigl(x_b-\bar X_b(t)\Bigr) f_i(x,t) dx,\\
&&
\bar X_a(t)=\int x_a f_i(x,t)dx,
\end{eqnarray}
where the dependence of $\sigma^2_{ab}$ and $\bar X_a$ on $i$ should
be understood implicitly. It should be noted that
$\sigma^2_{ab}(t=t_0)=(1/4)\delta_{ab}$. The realistic time evolution
of $\sigma^2_{ab}$ can be obtained by Vlasov equation as
\widetext
\begin{equation}
\dot\sigma^2_{ab}(t)\equiv {d\over dt}\sigma^2_{ab}(t)
=\int \Bigl[
 \Bigl(\dot x_a-\dot{\bar X}_a(t)\Bigr)\Bigl(x_b-\bar X_b(t)\Bigr)
+\Bigl(x_a-\bar X_a(t)\Bigr)\Bigl(\dot x_b-\dot{\bar X}_b(t)\Bigr)
 \Bigr] f_i(x,t)dx,
\label{eq:diffusion}
\end{equation}
\narrowtext\noindent
where $\dot x$ is the solution of the classical equation of motion
with the Hamiltonian $h$ for the phase space point $x$,
\begin{equation}
\{\dot x_a\}
=\Bigl\{\sqrt\nu\,\dot{\vec r},\; \dot{\vec p}/2\hbar\sqrt\nu\Bigr\}
=\Bigl\{\sqrt\nu{\partial h\over\partial{\vec p}},\;
      -{\partial h\over\partial{\vec r}}\Big/2\hbar\sqrt\nu\Bigr\},
\end{equation}
and
\begin{equation}
\dot{\bar X}_a(t)=\int \dot x_a f_i(x,t)dx.
\end{equation}
We can calculate $\dot\sigma^2_{ab}(t_0)$ using the Monte Carlo
integration method or the test particle method as is usually done in
solving Vlasov equation for heavy ion collisions. Since
$\dot\sigma^2_{ab}(t_0)$ can be diagonalized by an orthogonal
transformation, we can assume without losing generality that
\begin{equation}
\dot\sigma^2_{ab}(t_0)=\dot\sigma^2_a\delta_{ab}.
\end{equation}
It can be easily proved that
\begin{equation}
\mathop{\rm Tr}[\dot\sigma^2_{ab}(t_0)]
=\sum_{a=1}^6 \dot\sigma^2_a = 0,
\end{equation}
for the Gaussian wave packet $f_i(x,t_0)$. This relation can be
considered as a representation of the Liouville theorem. In numerical
calculations we find that three of $\{\dot\sigma^2_a\}_{a=1,\ldots,6}$
are positive and three of them are negative in most cases. As we have
mentioned before, we cannot treat the shrinking components but we
respect the diffusing components by giving the stochastic
displacement. Assuming that the reproduction of the second moment
$\dot\sigma^2_a$ of the diffusion is the most important and the effect
of the higher moments is negligible, we take the distribution function
of the stochastic displacement $g(\xi)$ to have a deformed Gaussian
form as
\begin{eqnarray}
&& g(\xi)=\prod_{a; \dot\sigma^2_a>0}
       \sqrt{2\alpha_a/\pi}\,e^{-2\alpha_a\xi_a^2}
       \prod_{a; \dot\sigma^2_a\le0} \delta(\xi_a),\\
&& \alpha_a=s^2/\dot\sigma^2_a,\qquad
   s^2={1\over6}\sum_{b=1}^{6}|\dot\sigma^2_b|,
\end{eqnarray}
where $s^2$ can be arbitrary but taken as above so that the typical
width of $g(\xi)$ is the same as that of the original wave packet
$f_i(x,t_0)$. Then it is easily proved that $\dot\sigma^2_{ab}(t_0)$
given by Vlasov equation are reproduced exactly by the stochastic
displacement for the diffusing directions in phase space if we take
the probability
\begin{equation}
c=4s^2\delta t.
\end{equation}
The choice of $s^2$ is rather arbitrary because frequent small
fluctuations and rare large fluctuations give the same diffusion
effect $\dot\sigma^2$. Although it is also possible to take $c=1$ and
$\alpha_a=(4\dot\sigma^2_a\delta t)^{-1}$ for example, we take the
above choice because the numerical calculation is easy when the
probability $c$ is small.

One may naively think that the above treatment of only the second
moment of the diffusion by the Gaussian stochastic displacement were
similar to the fermionic molecular dynamics (FMD) {}\cite{FELDMEIER}
which treats the widths of wave packets as dynamical variables. It is
true only for the instantaneous time evolution of the one-body
distribution function. The essential point of AMD-V is that each
component of the diffused wave packet can propagate independently of
other components and therefore it is allowed for the branched wave
packets to evolve on completely different trajectories. This feature
is very important in the nucleon emission process, for example, where
the emitted (high-momentum) component and the low-momentum component
remaining in the nucleus evolve in their own ways, while in FMD these
two components influence each other in a complex way because of the
nonlinearity of the approximate treatment of FMD.

\subsubsection{Recovery of the conservation lows}

The simple-minded stochastic displacement explained above brings about
the violation of conservation lows such as the total momentum and the
energy, and therefore the obtained final states cannot be interpreted
as physically meaningful final channels of the reaction.

The problem of the momentum conservation is rather trivial. By the
chain clustering method, we first define the nucleus that includes the
nucleon $i$ to which the stochastic displacement is to be given. Any
pair of two nucleons $j$ and $k$ with $|{\vec W}_j-{\vec W}_k| <1.25$
is considered to belong to the same nucleus and the mass number of the
nucleus which includes the nucleon $i$ is denoted by $A_{\rm nuc}$,
excluding the nucleon $i$. The conservation low of the total momentum
of this nucleus means that the fluctuation of the single-particle
momentum of the nucleon $i$ should be compensated coherently by the
other nucleons in this nucleus, though this correlation cannot be
described by the single-particle models. Therefore, to respect the
momentum conservation, we give the stochastic displacement $\xi$ not
to ${\vec W}_i$ but to the relative coordinate
\begin{equation}
{\vec W}_i-{1\over A_{\rm nuc}}\sum_{j\in{\rm nuc}}{\vec W}_j.
\end{equation}

The conservation of the total energy should be considered more
carefully. The branching by the stochastic process corresponds to the
decomposition of a Slater determinant to a superposition of Slater
determinants in the truly quantum mechanical description. Due to the
interference among the Slater determinants, they need not to be the
eigenstates of the energy in order for the total wave function to have
the definite energy. However, while they are evolving in time toward
the final channels of the reaction, their energies become the same as
the initial energy, because the matrix elements of the Hamiltonian
between different final channels are vanishing. In other words, the
energy deviation of each Slater determinant produced by the stochastic
displacement of a nucleon wave packet will be compensated by other
degrees of freedom of the nucleus before the system reaches the final
state. The standpoint of AMD-V is to respect the fluctuation and the
independence among the channels at the cost of the interference among
them. We neglect the finite time to recover the energy conservation
and require that the energy should be conserved just after the
stochastic displacement by adjusting other degrees of freedom of the
nucleus. Now the problem is how to decide the way of the energy
adjustment. The most natural requirement is that the energy
conservation should be achieved with the least modification of the
internal (canonical) coordinates of $A_{\rm nuc}$ nucleons in the
nucleus that includes the nucleon $i$ to which the stochastic
displacement has been given now. For this purpose, we solve the
constrained cooling/heating equation {}\cite{KANADAENYOa}
\begin{equation}
\mp\hbar\sum_{j\tau}C_{k\sigma,j\tau}{dZ_{j\tau}\over d\beta}
={\partial{\cal H}\over\partial Z^*_{k\sigma}}
+\sum_{l=1}^{n}\eta_l{\partial G_l\over\partial Z^*_{k\sigma}},
\label{eq:cooling}
\end{equation}
until the energy becomes equal to the initial value with a reasonable
precision. Here we have introduced several real functions
\(
\{G_l(Z)\}_{l=1,\ldots,n}
\)
of the constraints which include the displaced coordinate ${\vec W}_i$
of the nucleon $i$, the center-of-mass coordinate of $A_{\rm nuc}$
nucleons
\(
\sum_{j\in{\rm nuc}}{\vec W}_j/A_{\rm nuc}
\),
and the coordinates of irrelevant nucleons which do not belong to the
nucleus that includes the nucleon $i$. The Lagrange multipliers
$\{\eta_l\}$ should be determined so that the constraints are kept,
$dG_l/d\beta=0$. We further require that the energy adjustment should
not change collective coordinates and include following constraint
functions into $\{G_l\}$,
\begin{equation}
{\langle\Phi(Z)|\sum_k{\vec r}_k\times{\vec p}_k|\Phi(Z)\rangle
 \over\langle\Phi(Z)|\Phi(Z)\rangle}
\end{equation}
and
\begin{equation}
{\langle\Phi(Z)|\sum_k r_{k\sigma} r_{k\tau}|\Phi(Z)\rangle
 \over\langle\Phi(Z)|\Phi(Z)\rangle},\quad
{\langle\Phi(Z)|\sum_k p_{k\sigma} p_{k\tau}|\Phi(Z)\rangle
 \over\langle\Phi(Z)|\Phi(Z)\rangle},
\end{equation}
with $\sigma,\tau=x,y,z$. These constraints of collective coordinates
are based on the idea that the collective energy, such as the incident
energy in heavy ion collisions, should not be converted directly to
the energy for the stochastic displacement which arises from the
independent single-particle motions.

One might have the opinion that it is not necessary to conserve the
total energy because the AMD wave function $\Phi(Z)$ is not an energy
eigenstate. However, in the calculation of the nuclear reactions, the
initial state is almost an energy eigenstate because AMD can describe
the ground states of nuclei very well. As a matter of course, this
small energy dispersion should be kept constant through the reaction
in the exact solution. The reason why the AMD states $\Phi(Z)$ during
the reaction have large energy dispersion is not because it is
physical situation but because the AMD functional space is limited so
that all (thermally) excited states in AMD inevitably have large
energy dispersion. Therefore it is not appropriate to take the energy
dispersion of the AMD state as the physical one. It is rather
reasonable to neglect this spurious energy dispersion in the procedure
for the recovery of the energy conservation as explained above.

\subsubsection{Treatment of exceptional situations}

There are two kinds of exceptional (but frequently happening)
situations in which the procedures described above do not go straight
and the special care is required.

The first possibility is that the state $W$ after the stochastic
displacement is Pauli-forbidden {}\cite{ONOab} and there is no
corresponding AMD wave function $\Phi(Z)$. Should we cancel this
stochastic displacement like the Pauli-blocking in the two-nucleon
collision process, or should we try again by generating another random
number for the displacement? In order to answer this question, we
first note that Vlasov equation already respects the Pauli principle
in a semiclassical manner because the Liouville theorem in the
classical dynamics ensures that $f({\vec r},{\vec p},t)\le 1$ is
satisfied for any $t$ if the initial state satisfies the semiclassical
Pauli principle $f({\vec r},{\vec p},t_0)\le 1$ for each spin-isospin
state. Since the stochastic displacement has been decided by the time
evolution of $f$ according to Vlasov equation, no further
consideration of the Pauli principle is necessary. The state $W$ after
the stochastic displacement can be Pauli-forbidden because of the
mismatch between the exact quantum treatment of the Pauli principle in
AMD and the approximate semiclassical treatment in Vlasov equation.
If we simply canceled the stochastic displacement, the correct
diffusion given by Eq.\ (\ref{eq:diffusion}) would not be
obtained. Therefore, in the case of Pauli-forbidden $W$, we should try
again by generating another random number for the stochastic
displacement.

The second possibility is that the energy conservation is not achieved
by solving Eq.\ (\ref{eq:cooling}) even when the system has cooled
down to the energy minimum state under the given constraints. Should
we cancel this stochastic process or try again by generating another
value of the stochastic displacement?  This situation often happens in
the nucleus with low excitation energy and always in the ground state. 
One of the origin of this case is the mismatch between the quantum
mechanics in AMD (or in TDHF) and the semiclassical treatment in
Vlasov equation. The quantum one-body distribution function in AMD (or
TDHF) has high-momentum component even for the bound states such as
the ground state. Since we put this one-body distribution function
into Vlasov equation as the initial state at $t=t_0$, the
high-momentum component begins to go out of the nucleus and
contributes to the diffusion calculated by Eq.\ (\ref{eq:diffusion}),
which should not happen if we treat quantum mechanically.  Another
origin of the impossible energy recovery is the mismatch between the
many-body treatment in AMD and the averaged one-body treatment in
Vlasov (or TDHF) equation. For example, if the excitation energy of
the nucleus is less than the nucleon separation energy, no nucleon
emission is possible in AMD due to the energy conservation. However,
in Vlasov (or TDHF) equation, some part of a nucleon can go out
without violating the conservation of the averaged energy. Since the
energy of each branch (with or without nucleon emission) should be the
same as the initial energy, the result of Vlasov (or TDHF) equation is
unsatisfactory. From these consideration, we can say that the
diffusion of the wave packet that cause the impossible energy recovery
is spurious and should not have been included in Eq.\
(\ref{eq:diffusion}) based on which this stochastic displacement is
being considered. Therefore we should cancel this stochastic
displacement rather than try again.

In summary, in order to fix the mismatches between AMD and Vlasov
equation, the probability $c$ and the distribution function $g(\xi)$
of the stochastic displacement have been modified to $c'$ and
$g'(\xi)$,
\begin{eqnarray}
&& c'=c\;{\displaystyle
   \int\Theta_{\rm E}(X_i+\eta)\Theta_{\rm P}(X_i+\eta)g(\eta)d\eta
   \over\displaystyle
   \int\Theta_{\rm P}(X_i+\eta)g(\eta)d\eta},\\
&& g'(\xi)={\Theta_{\rm E}(X_i+\xi)\Theta_{\rm P}(X_i+\xi)g(\xi)
   \over\displaystyle
   \int\Theta_{\rm E}(X_i+\eta)\Theta_{\rm P}(X_i+\eta)g(\eta)d\eta},
\end{eqnarray}
with step functions of phase space point $x$,
\begin{eqnarray}
&& \Theta_{\rm P}(x)=\Biggl\{
          \begin{array}{ll}
          1 & \mbox{if $x$ is Pauli-allowed,} \\
          0 & \mbox{if $x$ is Pauli-forbidden.}
          \end{array}\\
&& \Theta_{\rm E}(x)=\Biggl\{
          \begin{array}{ll}
          1 & \mbox{if energy adjustment is possible,} \\
          0 & \mbox{if energy adjustment is impossible.}
          \end{array}
\end{eqnarray}

\section{Application to heavy ion collisions}
\widetext
\begin{figure}
\ifx\epsfbox\undefined\else
\begin{center}
\begin{minipage}{0.9\textwidth}
\epsfxsize\textwidth\epsfbox{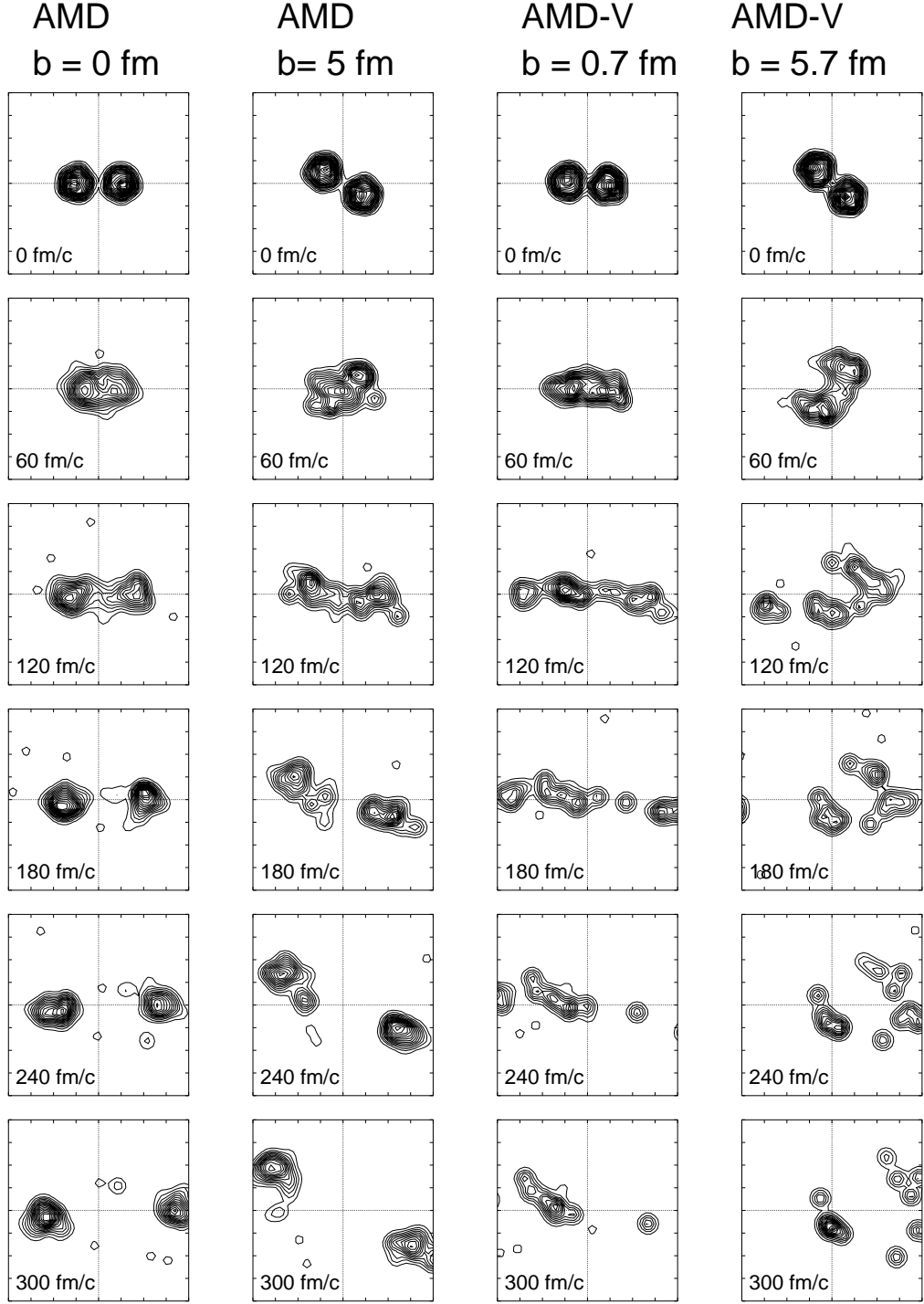}
\end{minipage}
\end{center}
\fi
\caption{\label{fig:animp}
Examples of the time evolution of the density projected onto the
reaction plane from $t=0$ fm/$c$ to $t=300$ fm/$c$ for
$\calcium+\calcium$ collisions at 35 MeV/nucleon. The size of the
shown area is $40\,{\rm fm}\times 40\,{\rm fm}$. Calculated results
with AMD (left two columns) and with AMD-V (right two columns) are
shown for impact parameters $b\sim 0$ fm and $b\sim5$ fm.}
\end{figure}
\narrowtext

In this section, we will show the calculated results of
$\calcium+\calcium$ reaction at 35 MeV/nucleon with AMD and AMD-V in
order to demonstrate the importance of the newly introduced stochastic
process of AMD-V to take account of the diffusion and the splitting of
wave packets according to Vlasov equation.

In Ref.\ {}\cite{HAGELb}, this reaction system was analyzed with
various models, including the quantum molecular dynamics (QMD)
{}\cite{MARUb}, the statistical sequential decay model ({\sc gemini})
{}\cite{GEMINI}, and the simultaneous multifragmentation model of
Gross {}\cite{GROSS}.  None of these models reproduces the $\alpha$
particle multiplicity $M_\alpha$ which is as large as 5.2. Most models
underestimate $M_\alpha$ by a factor more than 2, while they
overestimate the proton multiplicity $M_p$ which is 6.2 in experiment. 
Although the microscopic dynamical calculation of QMD gives better
result of the charge distribution of intermediate mass fragments
(IMFs) than the other statistical models, the unsatisfactory point is
that the QMD dynamical calculation should be truncated early at
$t\sim100$ fm/$c$ and should be connected to a statistical decay model
such as {\sc gemini} in order to get good results. This means that the
microscopic dynamical QMD calculation cannot describe the whole stage
of the reaction consistently. It should be commented here that in the
QMD used in Ref.\ {}\cite{HAGELb} the kinetic energy does not include
the zero-point oscillation energy though the density is the sum of
Gaussian wave packets, and hence the extension formulated in the
previous section is not directly applicable to this QMD.

\widetext
\begin{figure}
\begin{minipage}[t]{0.5\textwidth}
\ifx\epsfbox\undefined\else
\epsfxsize0.9\textwidth\epsfbox{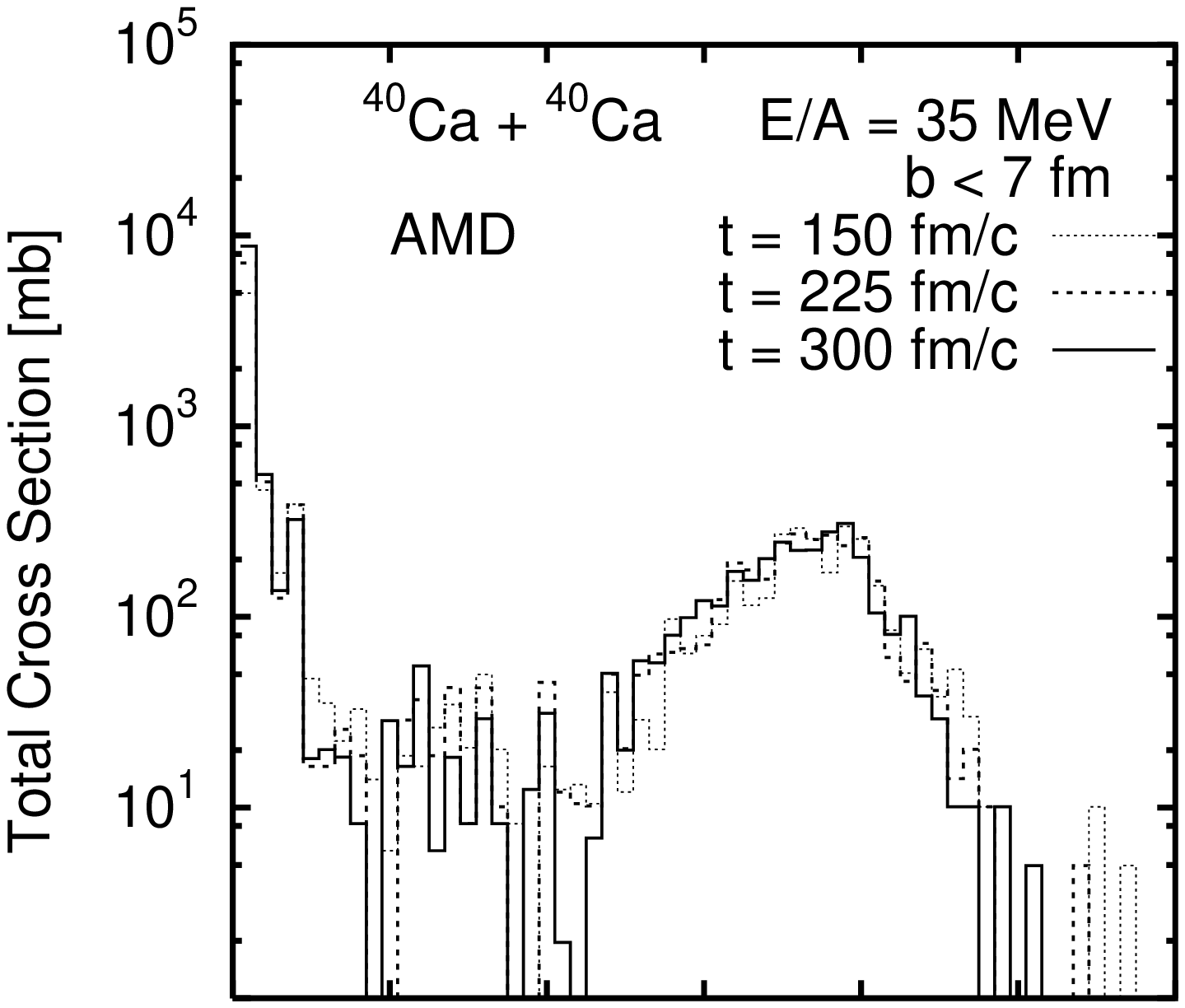}
\vspace{-0.219\textwidth}
\epsfxsize0.9\textwidth\epsfbox{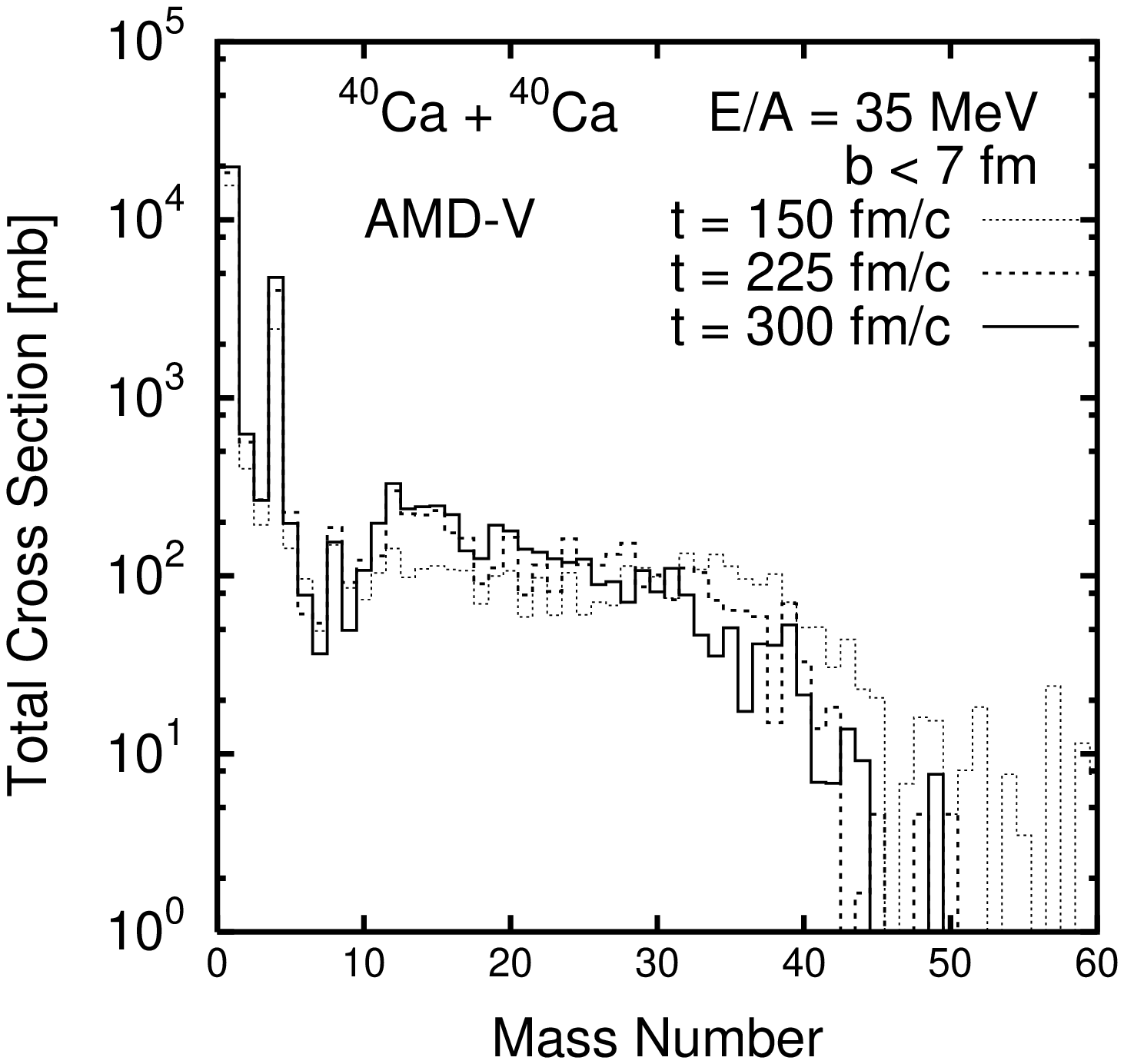}
\fi
\begin{minipage}{0.9\textwidth}
\caption{\label{fig:msdst-dy}
Mass distribution of fragments that exist at $t=150$, 225, and 300
fm/$c$ in the dynamical calculations with AMD (upper part) and AMD-V
(lower part).  }
\end{minipage}
\end{minipage}
\hspace{-0.02\textwidth}
\begin{minipage}[t]{0.5\textwidth}
\ifx\epsfbox\undefined\else
\epsfxsize0.9\textwidth\epsfbox{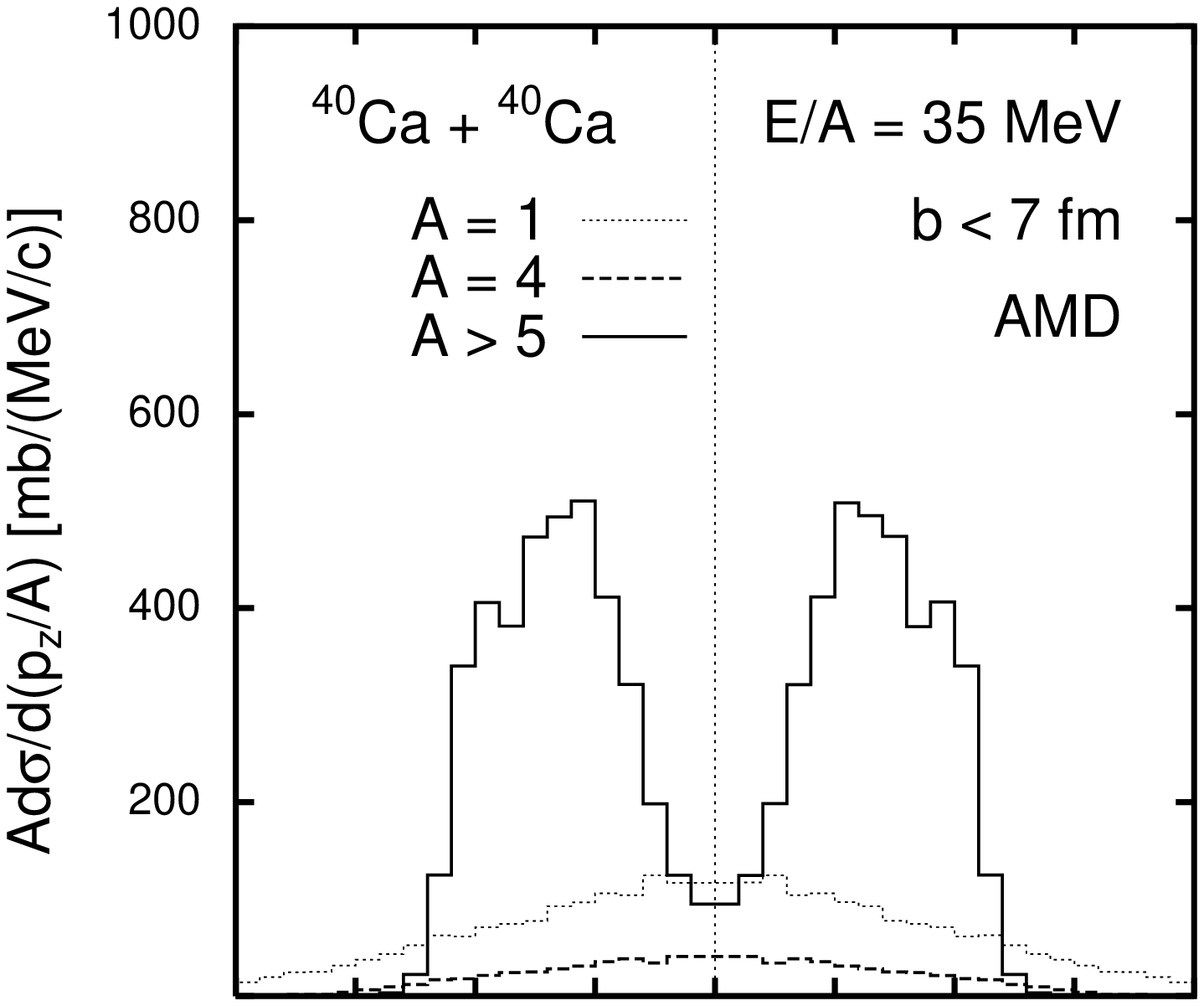}
\vspace{-0.219\textwidth}
\epsfxsize0.9\textwidth\epsfbox{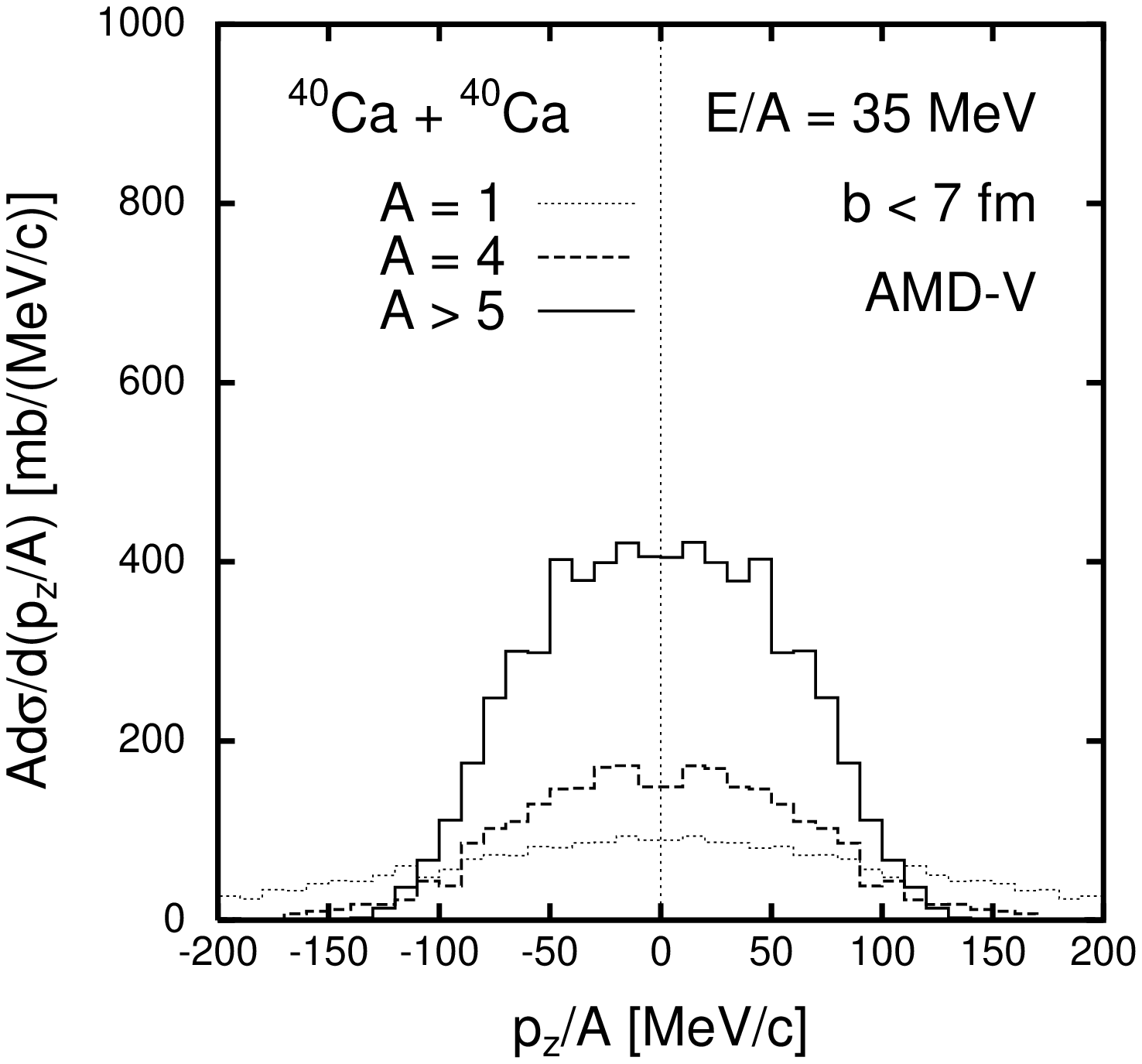}
\fi
\begin{minipage}{0.9\textwidth}
\caption{\label{fig:pzdst}
Parallel momentum distribution of nucleons, $\alpha$ particles and
heavier fragments after the statistical decay calculated with AMD
(upper part) and AMD-V (lower part). The vertical scale is
proportional to the number of nucleons contained in the
fragments. Incident momenta of the projectile and the target
correspond to $p_z/A=\pm 128$ MeV/$c$.}
\end{minipage}
\end{minipage}
\end{figure}
\narrowtext

The calculations with AMD and AMD-V are performed as usual. Gogny
force {}\cite{GOGNY} is adopted as the effective interaction. Coulomb
force is included. The width parameter is taken to be $\nu=0.16$ ${\rm
fm}^{-2}$. Two-nucleon collision cross section is the same as that
used in Ref.\ {}\cite{ONOe}. It should be noted that there is no
parameter introduced for the new stochastic process of AMD-V. Many
simulations are repeated because each simulation corresponds to an
experimental event. The impact parameter region $b<7$ fm is
investigated in this paper. Each dynamical simulation is continued
until $t=300$ fm/$c$. The statistical decay of the excited fragments
which exist at $t=300$ fm/$c$ is calculated with a code {}\cite{MARUb}
which is similar to {\sc cascade} by P\"uhlhofer
{}\cite{PUHLHOFER}. Although it is originally an evaporation
calculation code, we take account of the sequential binary decay of a
parent nucleus into two daughter nuclei both of which may be
excited. The excitation energy of the lighter daughter nucleus is
limited to $E^*<40$ MeV.

\begin{figure}
\ifx\epsfbox\undefined\else
\begin{minipage}[b]{0.5\textwidth}
\epsfxsize0.9\textwidth\epsfbox{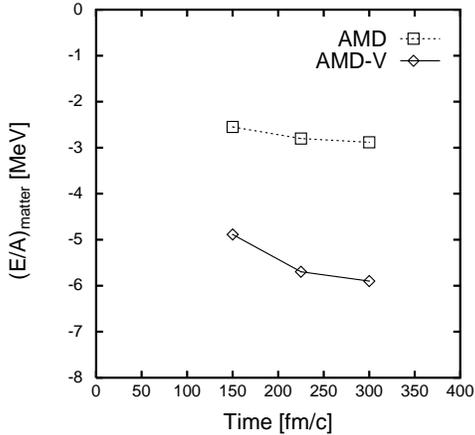}
\end{minipage}
\fi
\begin{minipage}[b]{0.48\textwidth}
\caption{\label{fig:deexeng}
Internal energy per nucleon of the matter part of the system (see
text) in events with $b<5$ fm as a function of the time in the
dynamical calculation of AMD (squares) and AMD-V (diamonds).  }
\end{minipage}
\end{figure}

\begin{figure}
\begin{minipage}[t]{0.5\textwidth}
\ifx\epsfbox\undefined\else
\epsfxsize0.9\textwidth\epsfbox{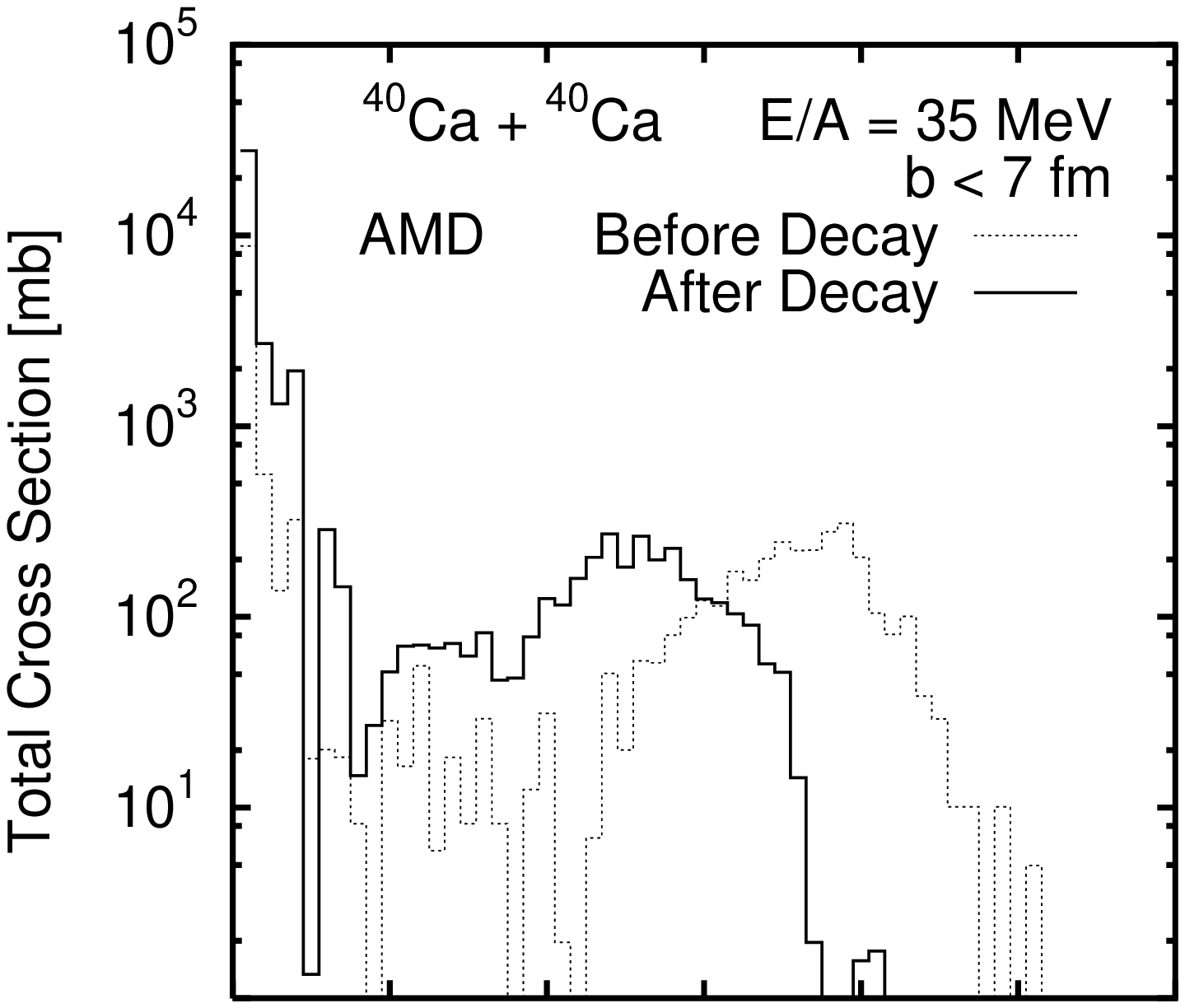}
\vspace{-0.219\textwidth}
\epsfxsize0.9\textwidth\epsfbox{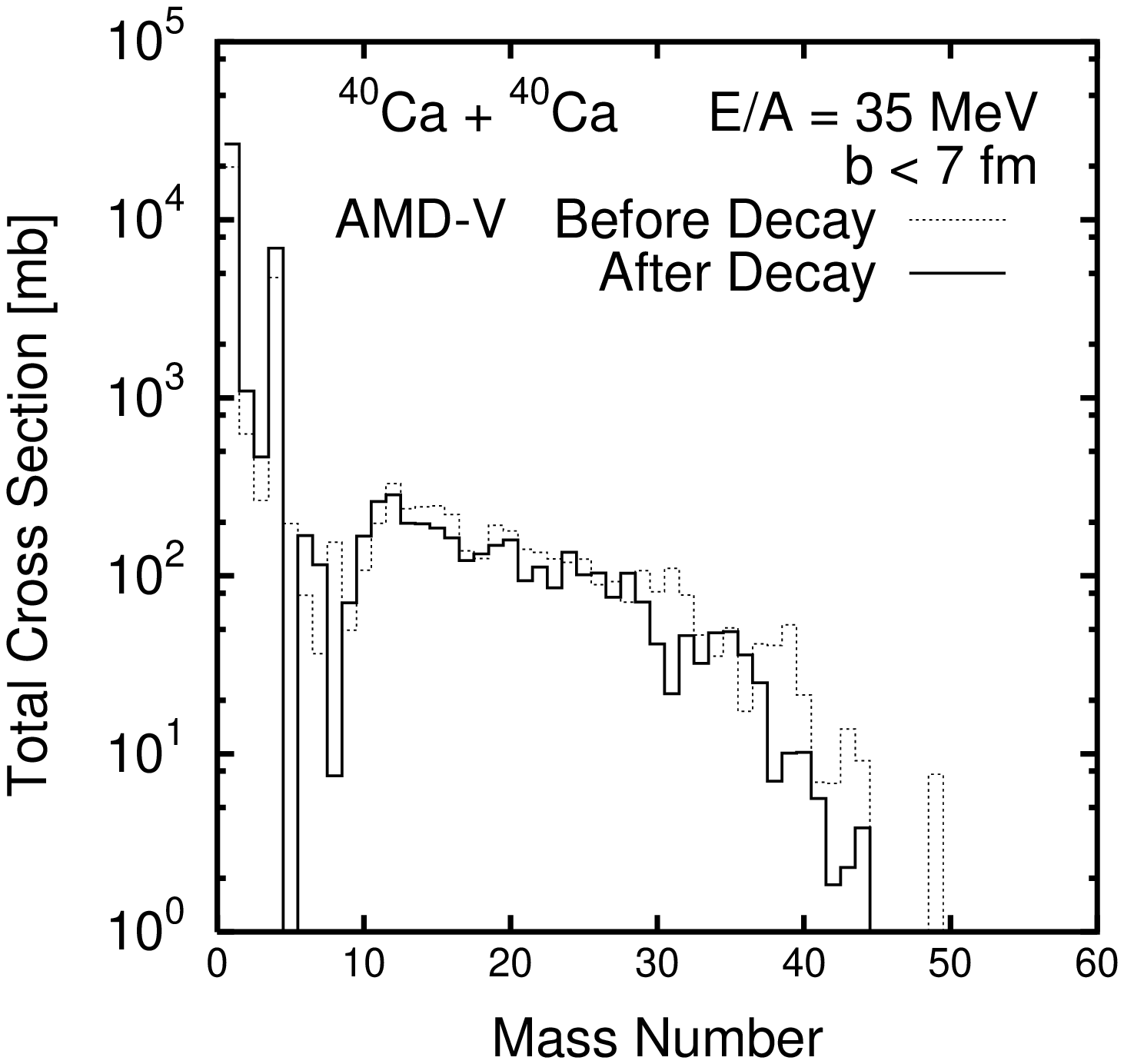}
\fi
\begin{minipage}{0.9\textwidth}
\caption{\label{fig:msdst}
Fragment mass distribution at the end of the dynamical calculation
(dotted histogram) and after the statistical decay calculation (solid
histogram). Results of AMD (upper part) and AMD-V (lower part) are
shown.}
\end{minipage}
\end{minipage}
\hspace{-0.02\textwidth}
\begin{minipage}[t]{0.5\textwidth}
\ifx\epsfbox\undefined\else
\epsfxsize0.9\textwidth\epsfbox{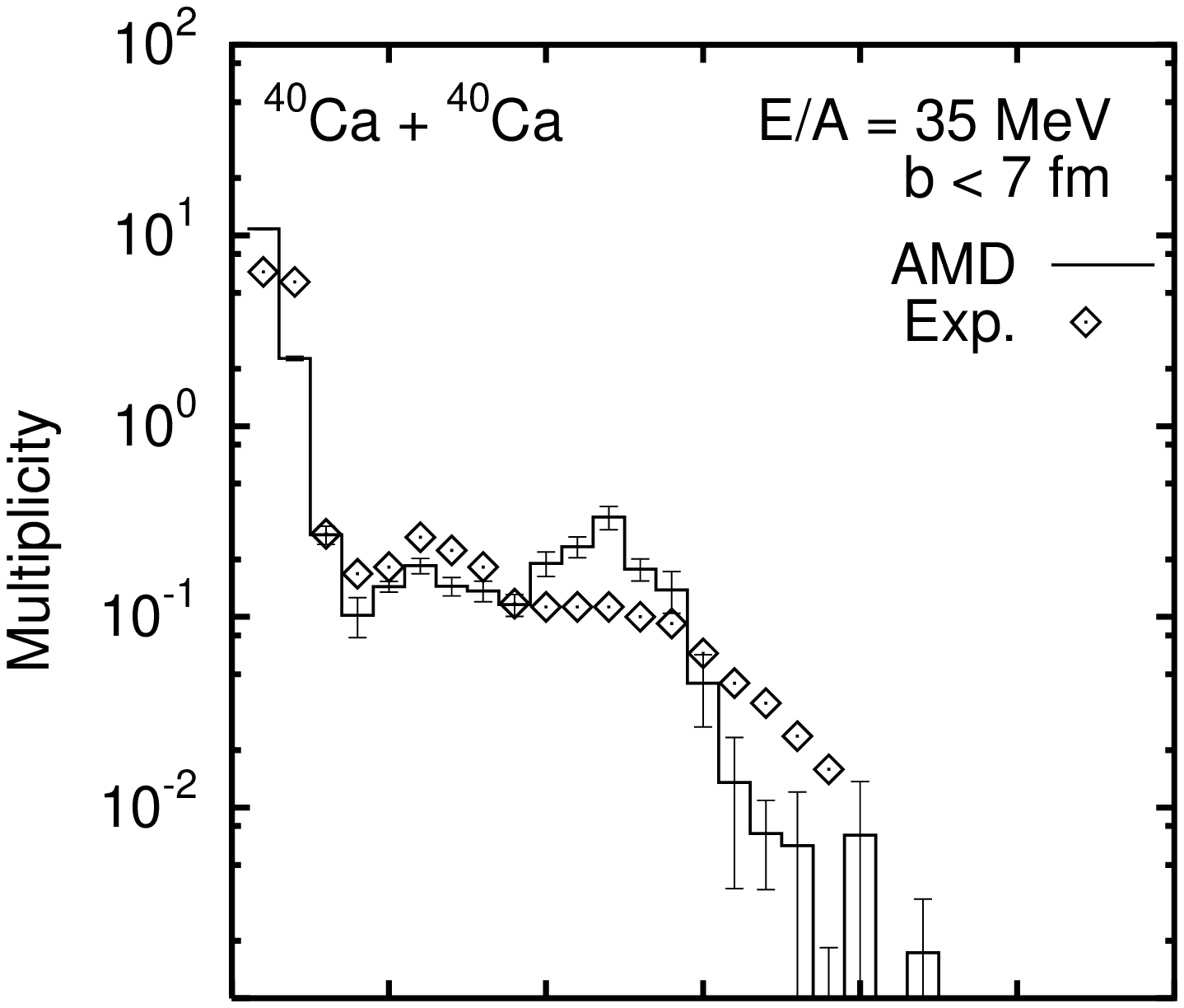}
\vspace{-0.219\textwidth}
\epsfxsize0.9\textwidth\epsfbox{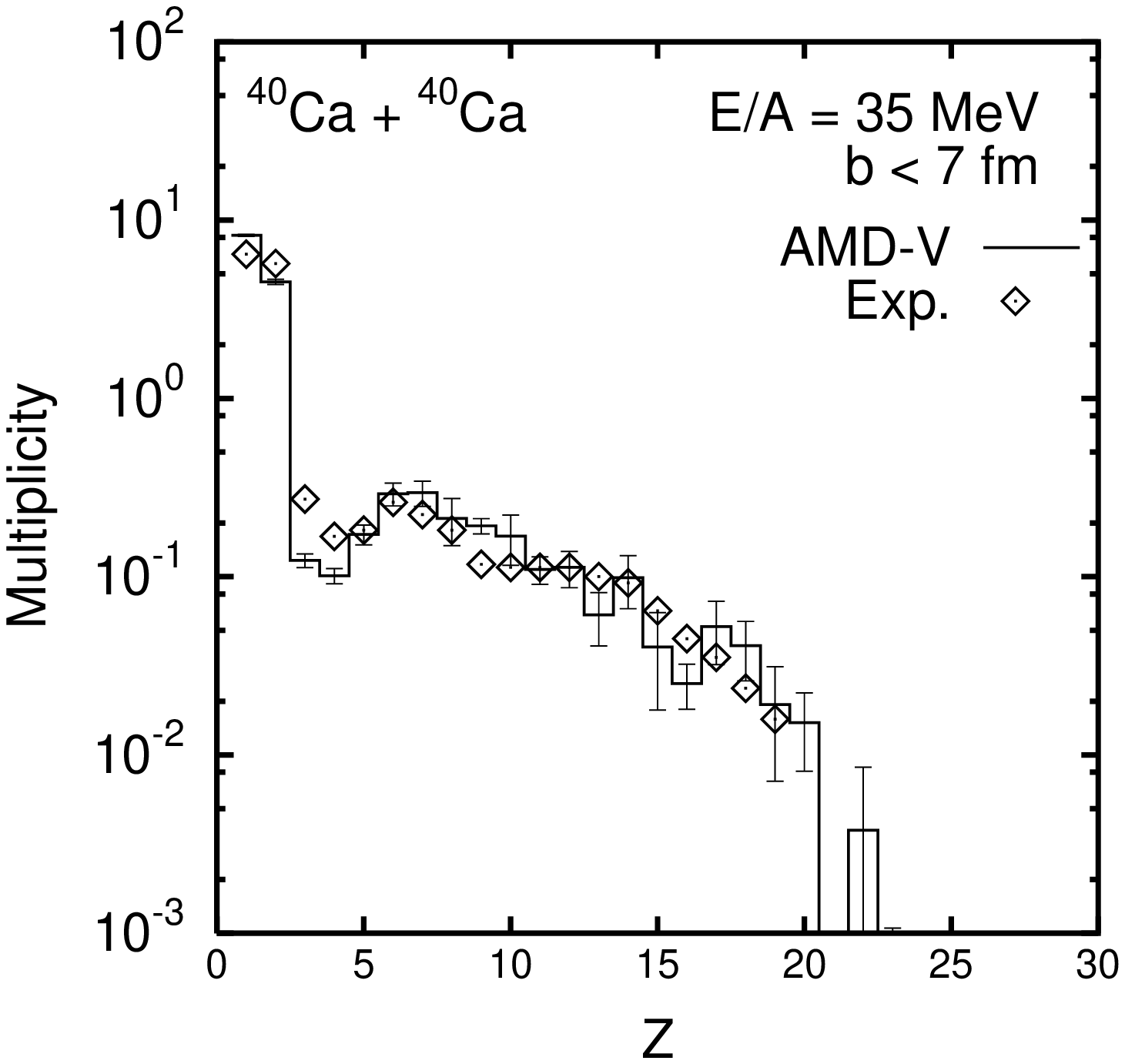}
\fi
\begin{minipage}{0.9\textwidth}
\caption{\label{fig:zmulti}
Calculated charge distribution (histogram) compared with the
experimental data (diamonds). The experimental filter has been
applied. Error bars show the estimated statistical error of the
calculated results. Results of AMD (upper part) and AMD-V (lower part)
are shown. }
\end{minipage}
\end{minipage}
\end{figure}

We first describe the features of the calculated results of the usual
AMD. Left two columns of Fig.\ {}\ref{fig:animp} show examples of the
time evolution of the density projected onto the reaction plane. We
can see that the projectile and the target pass through each other for
both impact parameters $b=0$ fm and $b=5$ fm. This is generally true
not only for these two events as can be seen in the upper part of
Fig.\ {}\ref{fig:msdst-dy} which shows the mass distribution of the
fragments that exist at $t=150$, 225, and 300 fm/$c$. There is a large
peak around $30\alt A\alt 40$ which is contributed by the
projectilelike and targetlike fragments in binary events. At $t=150$
fm/$c$, the system has already been separated into two large fragments
and the later time dependence of the mass distribution is rather
gradual. There is almost no yield for light IMFs with $10\alt A\alt
20$ in dynamical AMD calculation. The binary feature can be found not
only in the mass distribution but also in the momentum distribution of
fragments. Figure {}\ref{fig:pzdst} shows the distribution of the
momentum component parallel to the beam direction for particles with
$A=1$, $A=4$, and $A>5$ after the statistical decay calculation. For
fragments with $A>5$, the projectilelike and targetlike components are
clearly separated into two peaks which have been largely shifted and
dissipated from the incident values. Figure {}\ref{fig:deexeng} shows
the time-dependence of the internal energy of the matter part of the
system which is defined as
\begin{equation}
(E/A)_{\rm matter}=\sum_{k; A_k\ge 5} E_k
             \bigg/\sum_{k; A_k\ge 5} A_k,
\end{equation}
where $k$ is the label of the fragments that exist at the given time
$t$ in all events. $A_k$ and $E_k$ are the mass number and the
internal energy of the fragment respectively, and the sum is taken for
fragments with $A_k\ge 5$. As shown in the figure, the excitation
energy of the two large fragments produced in AMD calculation is very
high. If the ground state energy is assumed to be $-7$ or $-8$
MeV/nucleon, the averaged excitation energy of the fragments is about
5 MeV/nucleon. Furthermore, in spite of this very high excitation, the
cooling of the fragments is very slow. It will take a time of order of
1000 fm/$c$ for the fragments to lose 1 MeV/nucleon of the excitation
energy if the AMD calculation is continued for a long time. When the
decay of these fragments is calculated by the statistical decay code,
they emit many nucleons and other light particles and the mass
distribution of fragments changes largely as shown in the upper part
of Fig.\ {}\ref{fig:msdst}. The peak of the projectilelike and
targetlike fragments has shifted to left by about 15. Reflecting the
mass distribution before the statistical decay, there is a dip in the
IMF region of $10\alt A\alt20$. The yields of $\alpha$ particles and
light IMFs have increased very much by the statistical decay process
as well as the nucleon yield. In the case of $\alpha$ particles, only
15 \% of the final yield is due to the $\alpha$ particles produced in
the AMD calculation which is continued until $t=300$ fm/$c$.  It
should be noted that the fall of the yield for $A\agt25$ is due to the
finite impact parameter range $b<7$ fm.

The features of the usual AMD calculation are summarized as follows:
Reaction is almost always binary, and produced fragments are highly
excited but their decay is very slow. Only few $\alpha$ particles and
light IMFs are produced in the dynamical stage of the reaction, though
they can be produced by the statistical decay process. We do not think
these features are realistic as explained below.

On the very slow deexcitation of fragments, we have already discussed
in Refs.\ {}\cite{ONOf,ONOg}. Namely the usual AMD should
underestimate the nucleon emission rate from a hot nucleus because the
momentum distribution of the nucleon wave packet is not duely
reflected in the nucleon emission dynamics. In AMD, a nucleon cannot
go out of the nucleus unless the wave packet centroid can go out of
the nucleus, even though there should be some probability of nucleon
emission due to the high-momentum tail of the wave packet. Even when a
nucleon is emitted from the nucleus, the kinetic energy carried out by
the emitted nucleon will be very small as also shown in Ref.\
{}\cite{ONOf} as the classical caloric curve. It should be noted that
the wave packet spread in AMD is a very important factor for the
nucleon emission because the energy spread due to the wave packet
spread is $3\hbar^2\nu/2M= 10$ MeV per wave packet. Therefore the
usual AMD should be severely underestimating the nucleon emission rate
from the excited fragments. This point can be improved by AMD-MF
proposed in Ref.\ {}\cite{ONOf}. In AMD-MF, a stochastic fluctuation
is given to the momentum centroid of a nucleon according to the
momentum width of the wave packet when it is going out of the nucleus. 
By this method, the wave packet near the nuclear surface is allowed to
split into several branches such as the high-momentum component and
the low momentum component, and the high-momentum component can go out
of the nucleus with a reasonable probability. This improvement is
already contained in AMD-V which is a much more general framework than
AMD-MF. According to Vlasov equation, the nucleon wave packet which is
attacking the nuclear surface should change its shape in phase space
because the high-momentum component begins to go out of the nucleus
while the low-momentum component is reflected by the nuclear
surface. Since this diffusion effect according to Vlasov equation is
treated by AMD-V exactly, the nucleon emission is described
reliablely. In other words, the stochastic displacement is given to
the wave packet centroid so that the wave packet is shifted to a
high-momentum and outer-located value with a reasonable probability
and then it can go out of the nucleus.

The diffusion and the splitting of the wave packet should also affect
the binary feature of the reaction. It should be noted that the wave
packet centroids in the projectile or the target distribute in rather
compact region of phase space compared to the nucleon distribution
which is calculated as the centroid distribution folded by the width
of each wave packet. Therefore, even when the projectile density and
the target density overlap each other, the centroid distributions do
not overlap in peripheral collisions. In the true solution, the
nucleon transfer and/or the formation of participant hot region may
happen in the overlapped region. However, these phenomena are
impossible in AMD because the centroids in the projectile are passing
far from the target in peripheral collisions, even though there should
be some probability of nucleon transfer due to the tail component of
the wave packet density distribution. Furthermore, the wave packet
centroids in the two nuclei are well separated not only in the
coordinate space but also in the momentum space, and therefore the
spurious binary feature appears also in central collisions. The
two-nucleon collisions which happen in AMD calculation do not help to
remove the spurious binary feature, because most nucleons that are
scattered by two-nucleon collisions are simply emitted as single
nucleons.

\widetext
\narrowtext

Now let us turn to the results of AMD-V. The results have been shown
in Figs.\ {}\ref{fig:animp}--{}\ref{fig:msdst} together with the
results of the usual AMD. With the introduction of the stochastic
branching process based on the wave packet diffusion by Vlasov
equation, the calculated results of the fragmentation have changed
drastically.  As can be seen in Fig.\ {}\ref{fig:animp}, the reaction
is not binary any more. The dissipation of the incident energy is
large but not complete (Fig.\ {}\ref{fig:pzdst}). Sometimes an
elongated part is formed connecting the projectilelike and the
targetlike parts in the intermediate stage of the reaction and a
fragment can be left around the center-of-mass position in many
events. This fragmentation mechanism seems to correspond to the
interpretation of the neck fragmentation from the experimental
analysis by Montoya et al. {}\cite{LYNCH} As shown in Fig.\
{}\ref{fig:msdst-dy}, the yield of light IMFs with $10\alt A\alt 20$
has increased drastically compared to the case of the usual AMD. Most
light IMFs are produced before $t=225$ fm/$c$. It should be also
noticed that the yield of $\alpha$ particles which are produced in the
dynamical calculation is more than 10 times as large as in the usual
AMD calculation. Dynamically emitted nucleons are twice as many as in
the usual AMD. Although the system has been partitioned into smaller
pieces, the produced fragments are much less excited as shown in Fig.\
{}\ref{fig:deexeng}. The excitation energy is only about 2 MeV/nucleon
when a fragment is produced, and its deexcitation is more rapid than
in the usual AMD. Since the excitation energies of the fragments at
the end of the AMD-V calculation are already small, the statistical
decay calculation has only a small effect on the mass distribution as
shown in Fig.\ {}\ref{fig:msdst}. It should be noted that the final
yield of nucleons is slightly smaller than in the usual AMD, while the
final yield of $\alpha$ particles is much larger reflecting the large
yield of the dynamically produced $\alpha$ particles. The final mass
distribution of IMFs is very different from that of the usual AMD.

In order to compare the calculated results by AMD and AMD-V with the
experimental data {}\cite{HAGELb}, it is necessary to apply the
experimental filter. In experiment, only the completely detected
violent events are used for the analysis by selecting the events with
the detected total charge $Z_{\rm tot}\ge34$ and the detected charged
particle multiplicity $M_{\rm c}\agt10$. In the code of the
experimental filter, the arrangement of the detectors and their
thresholds are taken into account. In Fig.\ {}\ref{fig:zmulti}, the
calculated charge distribution of fragments after the application of
the experimental filter is compared with the data. We can see that the
result of the usual AMD is not good while the result of AMD-V is quite
satisfactory.  However, the underestimation of the light IMF yield by
the usual AMD is not so severe as could be expected from the inclusive
mass distribution in Fig.\ {}\ref{fig:msdst}. This is because of the
experimental filter. Due to the high detector threshold, the
targetlike fragments in the binary events are seldom detected and
therefore most of the binary events are rejected by the experimental
filter. On the other hand, the small number of ternary events are
relatively favored by the experimental filter. Therefore the filtered
result of the usual AMD does not reflect the feature of the total
events correctly, but the binary feature is still left in the filtered
charge distribution which is in contradiction to the data. The
overestimation of the proton multiplicity $M_p$ and the
underestimation of $\alpha$ multiplicity $M_\alpha$ are common to
other model calculations reported in Ref.\ {}\cite{HAGELb}. On the
other hand, AMD-V reproduces the charge distribution of fragments with
$Z\ge5$ almost perfectly, which demonstrates strongly the ability of
AMD-V to describe the fragmentation.  Furthermore AMD-V has improved
the problem of $M_p$ and $M_\alpha$ so much that it almost reproduces
them. The AMD-V results are $M_p=7.6$ and $M_\alpha=4.4$ while the
data are $M_p=6.2$ and $M_\alpha=5.2$.

\widetext
\begin{figure}
\ifx\epsfbox\undefined\else
\epsfxsize\textwidth\epsfbox{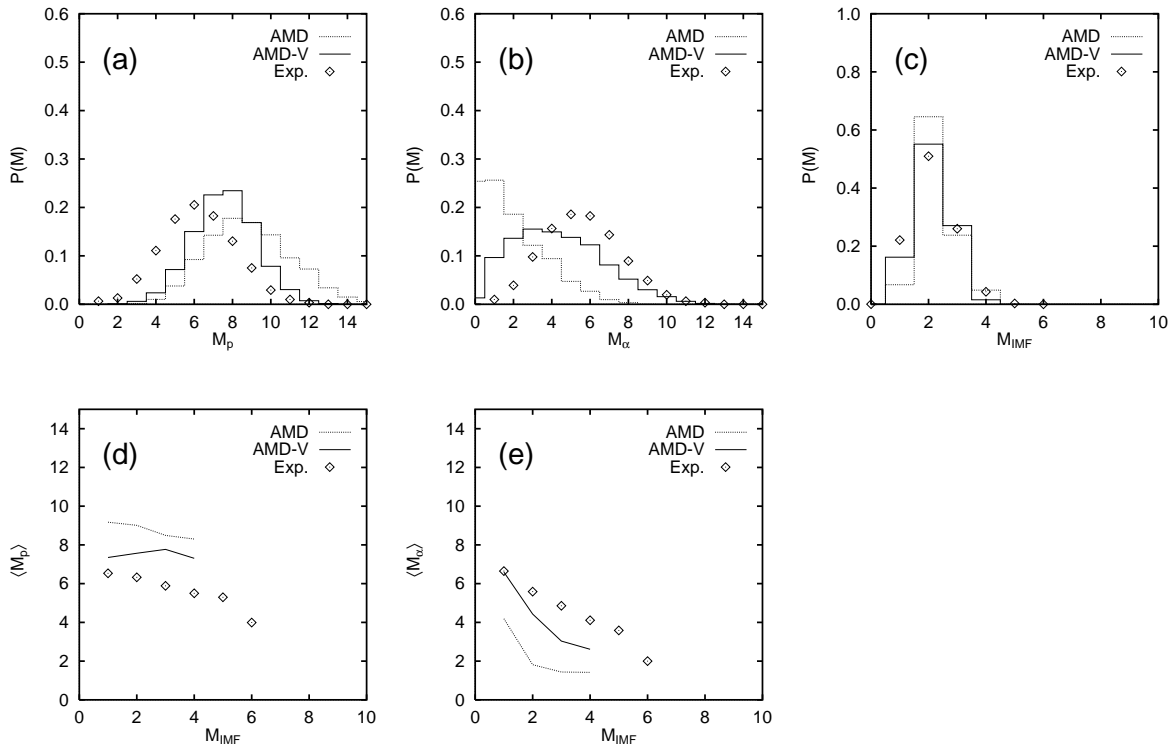}
\fi
\caption{\label{fig:multis}
Multiplicities of various products. Diamonds represent the
experimental data, dotted lines represent the results of AMD, and
solid lines represent the results of AMD-V. The experimental filter
has been applied. (a) Proton multiplicity, (b) $\alpha$ particle
multiplicity, (c) IMF multiplicity, (d) proton multiplicity vs IMF
multiplicity, and (e) $\alpha$ particle multiplicity vs IMF
multiplicity. In (d) and (e), calculated results are not shown for
$M_{\rm IMF}=0$ and $M_{\rm IMF}\ge5$ because the number of samples is
small.}
\end{figure}
\narrowtext

\begin{figure}
\ifx\epsfbox\undefined\else
\begin{minipage}[b]{0.5\textwidth}
\epsfxsize0.9\textwidth\epsfbox{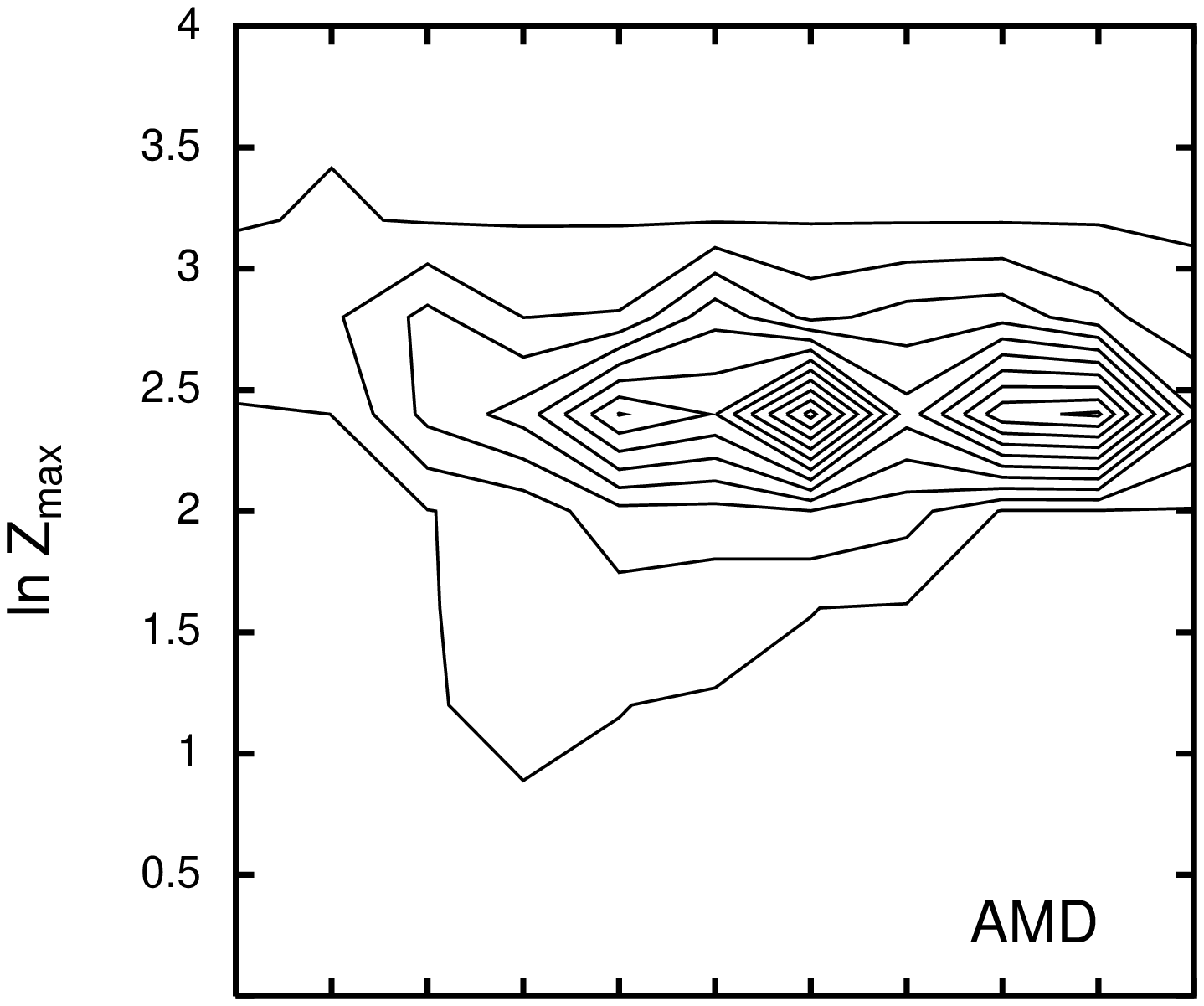}
\vspace{-0.219\textwidth}
\epsfxsize0.9\textwidth\epsfbox{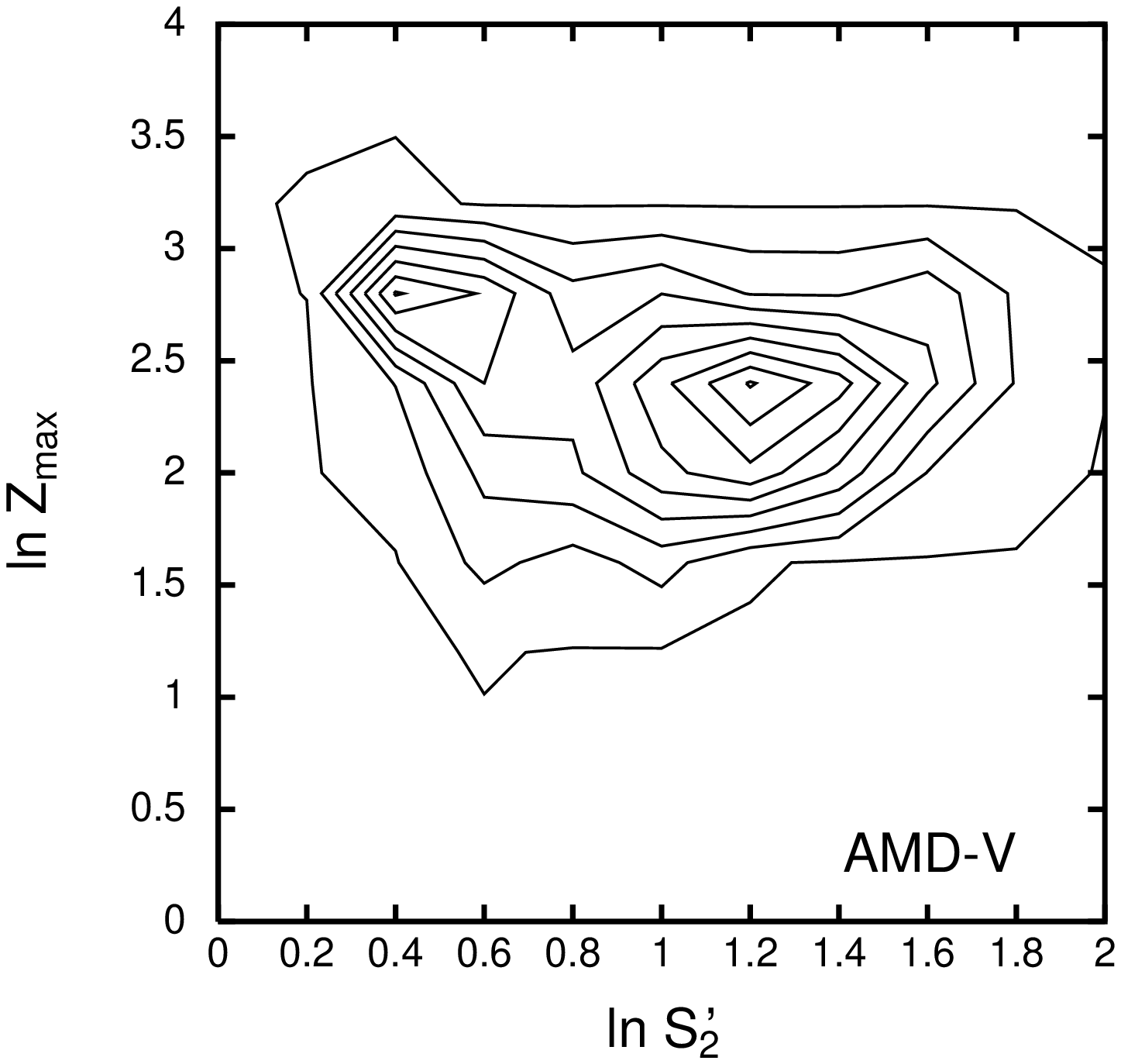}
\end{minipage}
\fi
\begin{minipage}[b]{0.48\textwidth}
\caption{\label{fig:s2dst}
Logarithmic distribution of $Z_{\rm max}$ vs $S_2'$ (see text). Each
contour represents constant value of $d^2P/d\ln S_2'd\ln Z_{\rm max}$
where $P$ is the probability normalized to 1. The outside contour is
at a level of 0.01, and each inner contour represents a progressive
increase of 0.15. Results of AMD (upper part) and AMD-V (lower part)
are shown after the application of the experimental filter.  }
\end{minipage}
\end{figure}

Figure {}\ref{fig:multis} shows the event-by-event analysis of the
fragmentation pattern, such as the probability distribution of $M_p$,
$M_\alpha$ and the IMF multiplicity $M_{\rm IMF}$. The event-by-event
correlation between $M_{\rm IMF}$ and $M_p$ and that between $M_{\rm
IMF}$ and $M_\alpha$ are also shown. The degree of the reproduction by
the usual AMD is similar to that by other models in Ref.\
{}\cite{HAGELb}, while the AMD-V result is much better than any other.
Although $M_p$ and $M_\alpha$ have been almost reproduced by AMD-V,
more $\alpha$ particles and less protons should be produced in events
with large $M_{\rm IMF}$ in order to get perfect results. We show in
Fig.\ {}\ref{fig:s2dst} the event-by-event distribution of the charge
$Z_{\rm max}$ of the largest fragment versus the normalized second
moment $S_2'$ of the event charge distribution with the largest
fragment excluded,
\begin{equation}
S_2'=\sum_{\!\!\!i; Z_i\ne Z_{\rm max}\!\!\!} Z_i^2
\bigg/ \sum_{\!\!\!i; Z_i\ne Z_{\rm max}\!\!\!} Z_i,
\end{equation}
where $i$ is the label of the fragments produced in an event. AMD-V
again reproduces very well the experimental data shown in Fig.\ 12 of
Ref.\ {}\cite{HAGELb} in which there are two components, one with small
$S_2'$ and large $Z_{\rm max}$ and the other with large $S_2'$ and
small $Z_{\rm max}$. The first component corresponds to the events
with a large IMF and many small fragments while the second component
is due to the events with several IMFs produced. On the other hand, in
the usual AMD result, the first component is missing because most
events are binary. Binary events have produced a peak with too large
$S_2'$.

\begin{figure}
\ifx\epsfbox\undefined\else
\begin{minipage}[b]{0.5\textwidth}
\epsfxsize0.9\textwidth\epsfbox{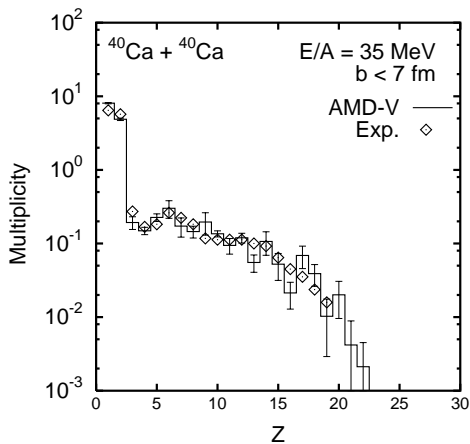}
\end{minipage}
\fi
\begin{minipage}[b]{0.48\textwidth}
\caption{\label{fig:zmulti-vx}
The same as the lower part of Fig.\ \protect\ref{fig:zmulti} but the
light fragments ($A\le13$) are put into the statistical decay
calculation immediately when they are produced after $t=150$ fm/$c$
unless they are absorbed again by a heavy fragment ($A\ge14$) before
$t=300$ fm/$c$.}
\end{minipage}
\end{figure}

The only unsatisfactory point of the AMD-V result in Fig.\
{}\ref{fig:zmulti} is the underestimation of the multiplicity for
$Z=3$ and 4. By investigating the time evolution of the system, we
have found that light IMFs with $5\alt A\alt 10$ are often produced
from time to time but they tend to merge with one another to form a
larger fragment again. If this merging is spurious, we would get very
good result by removing it, as shown in Fig.\
{}\ref{fig:zmulti-vx}. In getting this result, we have assumed that
the light IMFs with $A\le13$ are put into the statistical decay code
immediately after they are produced, unless they are absorbed by a
large fragment with $A\ge14$ in later stage of the dynamical
calculation. A possible reason why the merging of light IMFs may be
spurious is that the diffusion and the splitting of the center-of-mass
wave packets of clusters in a nucleus is not fully described in AMD-V
when they are emitted from a nucleus. In AMD-V, the dispersion and the
diffusion of the center-of-mass wave packet of a cluster is described
through those of the nucleons contained in the cluster and therefore
the diffusion of the center-of-mass wave packet is always accompanied
by the internal excitation of the cluster. If the center-of-mass
diffusion is treated separately from the internal degrees of freedom
as it should be, more energy will be available for the translational
kinetic energy of the cluster and the light IMF emission will
increase. It should be noted that the zero-point oscillation energy
per nucleon of the cluster center-of-mass motion is
$3\hbar^2\nu/2AM\approx10/A$ MeV with $A$ being the mass number of the
cluster, and therefore it is less important for heavier clusters.

\section{Relation of AMD-V to other models}

In this section, we intend to clarify the interrelation among the
frameworks of microscopic models which have been used for the study of
the dynamics of heavy ion collisions, such as TDHF, VUU, QMD, AMD, and
AMD-V. Since the nuclear reaction system is a quantum system and is
described by a time-dependent wave function, we mainly discuss on the
quantum models such as TDHF and AMD. However, most of the following
discussions can be applied also to semiclassical models.

Before the discussion on the meaning of the newly introduced process
of AMD-V, we will first mention on the difference of the treatments of
two-nucleon collisions in one-body transport models and in molecular
dynamics models. Let us take the initial state as a Slater determinant
\begin{equation}
|\Psi_0\rangle=\det[\psi_{01}\psi_{02}\cdots\psi_{0A}].
\label{eq:initSlater}
\end{equation}
In order to compare AMD with TDHF, we may take $|\Psi_0\rangle$ to
be an AMD wave function $|\Phi(Z_0)\rangle$.  Here we consider
$|\Psi_0\rangle$ as the state just before the first chance of
two-nucleon collision in a heavy ion reaction. If the first chance of
two-nucleon collision happens between the first and the second
nucleons, the wave function of these two nucleons changes as
\begin{equation}
\psi_{01}\psi_{02}\rightarrow 
\sum_{\alpha\beta}c_{\alpha\beta}\psi_1^{(\alpha)}\psi_2^{(\beta)},
\end{equation}
while other nucleon wave functions propagate as
$\psi_{0i}\rightarrow\psi_i$ ($i=3,\ldots,A$) obeying the TDHF
equation or some approximated equation. $\alpha$ and $\beta$ represent
the single-particle states and one of the pairs of $\alpha\beta$
corresponds to the state without two-nucleon collisions. Accordingly,
the initial Slater determinant has changed into a linear combination
of many Slater determinants
\begin{eqnarray}
|\Psi_0\rangle \rightarrow 
&&|\Psi\rangle
=\sum_{\alpha\beta}c_{\alpha\beta}|\Psi^{(\alpha\beta)}\rangle,
\\
&&|\Psi^{(\alpha\beta)}\rangle
=\det[\psi_1^{(\alpha)}\psi_2^{(\beta)}\psi_3\cdots\psi_A].
\end{eqnarray}
Further successive two-nucleon collisions may cause more branching
into Slater determinants.  Since $|\Psi\rangle$ is not a Slater
determinant any more, TDHF cannot be applied to it. However, it should
be noted that each component $|\Psi^{(\alpha\beta)}\rangle$ is a
Slater determinant and TDHF may be applied to it as a good
approximation. Each $\alpha\beta$ has its own mean field calculated
from $|\Psi^{(\alpha\beta)}\rangle$ and the single-particle wave
functions propagate under this mean field. In order to decide the time
evolution of a component $|\Psi^{(\alpha\beta)}\rangle$, no
information of other components should be necessary because of the
linearity of the quantum mechanics. However, what is done in one-body
transport models like VUU and the extended TDHF for two-nucleon
collisions strongly contradicts to this point of view. The extended
TDHF (or VUU) calculates the effect of the two-nucleon collision on
the one-body density matrix $\rho({\vec r},{\vec r}')$ which can be
always defined as
\begin{equation}
\rho({\vec r},{\vec r}')=\langle\Psi|\hat\rho|\Psi\rangle,\qquad
\hat\rho=\sum_i |{\vec r}\rangle\langle{\vec r}'|_i,
\end{equation}
though the system is not described by a single Slater determinant
after the two-nucleon collision. The future time evolution of
single-particle states is determined by the TDHF-like equation with
the mean field calculated from $\rho({\vec r},{\vec r}')$ as if it
were not for any many-body correlation and as if the two-body density
matrix could be given by an antisymmetrized product of two one-body
density matrices. This means that the single-particle wave functions
of all components $|\Psi^{(\alpha\beta)}\rangle$ are propagated by a
common mean field, which is in contradiction to the linearity of the
time-dependent Schr\"odinger equation. On the other hand, in AMD (and
many other molecular dynamics models), one of the components
$|\Psi^{(\alpha\beta)}\rangle$ is chosen stochastically as a channel
with the probability $|c_{\alpha\beta}|^2$ when a two-nucleon
collision has happened. Here $\alpha\beta$ may be considered as the
scattering angle and $c_{\alpha\beta}$ as the scattering
amplitude. Namely, in AMD, the two-nucleon collisions are treated as
the stochastic branchings into channels each of which is represented
by an AMD wave function. Therefore the mean fields of different
channels are different and the time evolutions of channels are
independent of one another, which is a quite plausible feature. This
is the reason why AMD is superior to the one-body transport models in
the description of medium energy heavy ion collisions where the
branching effect is more important than the flexibility of
single-particle wave functions. A possible drawback of AMD treatment
is that the effect of the two-nucleon collisions is treated
stochastically by giving up the description of the interference among
the components $|\Psi^{(\alpha\beta)}\rangle$ after two-nucleon
collisions.

The above-mentioned importance of the two-nucleon collision process as
a source of the branching into channels should have been well known,
and it is not directly related to the newly introduced process of
AMD-V. Here we consider phenomena in which the two-nucleon collision
process is not important, in order to clarify the relation and the
difference among TDHF, AMD, and AMD-V. Let us take the Slater
determinant $|\Psi_0\rangle$ of Eq.\ (\ref{eq:initSlater}) as the
initial state. For example, $|\Psi_0\rangle$ may be an excited
fragment produced in one of the channels of a medium energy heavy ion
collision with $A$ being the mass number of this fragment. It is
usually sufficient to assume that $|\Psi_0\rangle$ is equal to an AMD
wave function $|\Phi(Z_0)\rangle$ because AMD can describe the excited
fragments in good approximation. According to TDHF equation, the
single-particle wave functions will propagate for a short period,
and the many-body wave function will change as
\begin{equation}
|\Psi_0\rangle \rightarrow |\Psi\rangle
=\det[\psi_1\psi_2\cdots\psi_A].
\label{eq:Slater}
\end{equation}
Now let us consider the case in which the first nucleon is going out
of the nucleus with some probability while the others remain in
it. (It is trivial to generalize the following discussion to the case
in which all nucleons can be emitted.) Then the propagated
single-particle wave function of the first nucleon can be decomposed
into two (or more) parts as
\begin{equation}
\psi_1=\sum_\alpha c_\alpha \psi_1^{(\alpha)},
\label{eq:singleDecomp}
\end{equation}
where $\psi_1^{(1)}$ is spatially localized in the nucleus and
$\psi_1^{(2)}$ is out-going part of $\psi_1$. Accordingly, the Slater
determinant $|\Psi\rangle$ can be written as a linear combination of
two (or more) Slater determinants as
\begin{eqnarray}
&&|\Psi\rangle=\sum_\alpha c_\alpha |\Psi^{(\alpha)}\rangle,
\label{eq:multiSlater}\\
&&|\Psi^{(\alpha)}\rangle=\det[\psi_1^{(\alpha)}\psi_2\cdots\psi_A].
\end{eqnarray}

Then, in order to solve the further time evolution of $|\Psi\rangle$,
we can think of two different ways. The first way is to continue
solving TDHF equation for $|\Psi\rangle$ of Eq.\ (\ref{eq:Slater}) as
is widely done in TDHF calculation. The second way is to apply TDHF
equation to each Slater determinant $|\Psi^{(\alpha)}\rangle$ and
solve the time evolution of each channel independently. Of course,
these two ways are equivalent for the exact quantum mechanics due to
the principle of superposition. However, it is not true for the
approximated treatment like TDHF where the nonlinearity has been
introduced for the sake of the feasibility of calculation. The first
way corresponds to using the common mean field to all channels of Eq.\
(\ref{eq:multiSlater}), while the mean fields for
$|\Psi^{(\alpha)}\rangle$ are different from channel to channel in the
second way. Which way should we take for the best description of the
system? It should be noted that $|\Psi^{(1)}\rangle$ represents a
nucleus of mass number $A$ with all nucleons spatially localized in
the nucleus, while $|\Psi^{(2)}\rangle$ is equivalent to the product
of the state of a nucleus with mass number $A-1$ and the state of the
out-going nucleon. Therefore we can say, at least, that the second way
is reliable as long as the TDHF description of nuclei with mass
numbers $A$ and $A-1$ is reliable. On the other hand, in the first
way, the common mean field for all channels is made by fractional
number of nucleons between $A-1$ and $A$, and such mean field cannot
be a good one for the nucleus with mass number $A-1$ nor $A$. Although
this pathological nature of TDHF for nucleon emission may not be so
serious in practice, this example explains how the breakdown of the
usual TDHF begins to grow up into the nonsense description of the
multichannel final state in heavy ion collisions even when the
two-nucleon collision effect is not directly important. From this
consideration, we can adopt an approximation that, when
single-particle wave functions are spreading wide like in the case of
nucleon emission, a Slater determinant should be decomposed into a
linear combination of channel Slater determinants by decomposing
single-particle wave functions into localized components and then the
time evolutions of channel Slater determinants should be solved
separately to get reliable results in the global time scale.

Now let us consider the case of AMD.  It should be noted first that
the TDHF result $|\Psi\rangle$ of Eq.\ (\ref{eq:Slater}) is more
reliable than the AMD result
$|\Phi(Z_0)\rangle\rightarrow|\Phi(Z)\rangle$ for the time evolution
of a short period because the TDHF single-particle wave function is
more general than that of AMD. When the nucleon-emission channel
$|\Psi^{(2)}\rangle$ is a minor branch, the AMD equation of motion
will simply abandon this channel and $|\Phi(Z)\rangle$ will be almost
equivalent to $|\Psi^{(1)}\rangle$.  Although the loss of minor
channels is not satisfactory, the time evolution of the main channel
will be described rather well just as in the second way of the
previous paragraph because the pathological situation of TDHF will not
occur in AMD due to the compact single-particle wave functions.

The meaning of the newly introduced process of AMD-V is now quite
obvious. At first the system is represented by an AMD wave function
$|\Psi_0\rangle=|\Phi(Z_0)\rangle$ and its time evolution for a short
period is calculated with TDHF (or Vlasov) equation. Each
single-particle wave function may begin to spread. Then it is
decomposed as in Eq.\ (\ref{eq:singleDecomp}) with $\psi_1^{(\alpha)}$
being Gaussian wave packets. Accordingly, the many-body wave function
is decomposed as in Eq.\ (\ref{eq:multiSlater}) into many AMD wave
functions $|\Psi^{(\alpha)}\rangle$ in this case. One of these
channels $|\Psi^{(\alpha)}\rangle$ is chosen with the appropriate
probability and its future time evolution is solved just in the same
way as was done for $|\Psi_0\rangle$ without any influence from other
channels. Thus AMD-V respects the minor channels which are lost in
AMD. It should be emphasized that TDHF (or Vlasov) equation is always
applied to the instantaneous time evolution of an AMD wave function
with compact single-particle wave functions and therefore AMD-V is
free from the pathological nature of TDHF that would appear for a
Slater determinant which should be decomposed into channels. In
principle, the best way among the arbitrary ways to decompose a Slater
determinant is to decompose the Slater determinant into channel Slater
determinants so that the future time evolutions of all the channels
can be calculated most precisely in mean field approximation. The
important standpoint of AMD-V is therefore to take the AMD wave
functions as the channel wave functions to which TDHF equation can be
applied most safely.

It should be commented that in the above explanation of AMD-V we have
ignored the interference among channels
$|\Psi^{(\alpha)}\rangle$. Generally speaking, it is very difficult to
respect the interference because the time evolutions of all channels
should be calculated very precisely in the global time scale in order
to get meaningful results with interference. This difficulty is
avoided in AMD-V by introducing semiclassical treatments by using
Vlasov equation instead of TDHF equation and QMD-like representation
of the one-body distribution function instead of the AMD wave
function. In other words, we have replaced the many-body density
matrix of a pure state $|\Psi\rangle$ with a statistical ensemble of
channel wave functions $|\Psi^{(\alpha)}\rangle$ as
\begin{equation}
|\Psi\rangle\langle\Psi|
=\sum_{\alpha\beta}c_\alpha c_\beta^*
                  |\Psi^{(\alpha)}\rangle\langle\Psi^{(\beta)}|
\approx \sum_\alpha w_\alpha
                  |\Psi^{(\alpha)}\rangle\langle\Psi^{(\alpha)}|,
\label{eq:density-matrix}
\end{equation}
where the probability $w_\alpha$ of branching is decided so that
Vlasov equation is reproduced as much as possible. The energy recovery
procedure explained in Sec.\ II.B.3 ensures that the channels can be
physical when they have evolved into the final state. The limitation
of channel wave functions $|\Psi^{(\alpha)}\rangle$ to AMD wave
functions is essential in many senses. As already discussed, the mean
field approximation can be applied to the AMD wave functions more
safely than to general Slater determinants. Furthermore, AMD wave
functions agree with the intuitive concept of `events' that the system
is divided into moving fragments composed of integer number of
nucleons in each event. Since one usually does not think of an
observable that connects different events (or channels), the
interference terms in Eq.\ (\ref{eq:density-matrix}) are usually
irrelevant. Only for the internal motions in a fragment, the
interference may be important when one is interested in the precise
single-particle wave functions. It should be finally emphasized that
we still have many-body wave functions $|\Psi^{(\alpha)}\rangle$ which
have quantum mechanical information within channels, in spite of the
semiclassical treatment of branching introduced to respect the
independence among channels which is also an important quantum
mechanical feature.

\section{Summary}

In this paper, we have presented a new version of AMD with stochastic
incorporation of Vlasov equation. In this AMD-V, the diffusion of wave
packets according to Vlasov equation is treated as the stochastic
branching into events. Namely, in addition to the equation of motion
and two-nucleon collisions of the usual AMD, stochastic displacements
are given to the centroids of wave packets so as to reproduce the time
evolution of one-body distribution function predicted by Vlasov
equation. No important free parameters are introduced for this new
process of AMD-V. The mean fields are different from event to event,
and therefore the independence among the time evolutions of events (or
channels) is treated properly unlike the TDHF (or Vlasov) calculation. 
It should be noted that at any time of each event we have an AMD wave
function to which Vlasov equation is applied. The merit of this fact
is not only that the mean field can be calculated easily but also that
the mean field approximation can be adopted more safely than for a
general Slater determinant which may have to be treated as a linear
combination of several channels. AMD-V is a fully quantum mechanical
framework with channel wave functions except that the interference
among channels cannot be calculated. However, the extension of AMD-V
is not directly related to the exact antisymmetrization in AMD and
therefore this extension is applicable to other molecular dynamics
models with wave packets. Although we have introduced AMD-V as an
extension of AMD to remove the restriction on the shape of wave
packets, it is also possible to regard AMD-V as giving a scheme to
extend the TDHF (or Vlasov) calculation to go beyond a Slater
determinant by taking account of the branchings into channels and
respecting the independence among them.

The ability of AMD-V to describe the fragmentation in medium energy
heavy ion collisions has been demonstrated by the very good
reproduction of data for $\calcium+\calcium$ system at 35 MeV/nucleon,
which have never been reproduced by any other model. The new
stochastic process of AMD-V has enabled the description of minor
branching processes which are irrelevant to the two-nucleon collision
process, such as the evaporation of nucleons from a hot nucleus and
the nucleon transfer in heavy ion collisions. In the case of
$\calcium+\calcium$ system at 35 MeV/nucleon, the spurious binary
feature of the AMD result has been removed by AMD-V and the abundant
production of light IMFs and $\alpha$ particles is described very well
mainly as the result of the formation of neck region between
projectilelike and targetlike components which have passed through
each other with dissipation. Produced fragments are already cool and
the statistical decay calculation is not so important as in the case
of AMD.

The proposition of AMD-V has derived from our study of the statistical
property of AMD. The successful description of fragmentation in heavy
ion collisions can also be interpreted as due to the improvement of
the statistical property. Namely, the improvement of the caloric curve
to the quantum mechanical one is important to reduce the excitation
energy of fragments. Furthermore, the pressure (i.e., the force
necessary to keep the volume) of the hot nuclear matter has been
increased by the new process of AMD-V, which facilitates the expansion
of the hot system followed by the fragmentation. However, what should
be emphasized here is that such an improvement is just a
straightforward result of the improvement of the microscopic
single-particle dynamics to enable the diffusion of wave packets.

AMD-V adopts Vlasov equation for the instantaneous time evolution of
the AMD wave function. Although this mean field approximation seems to
have given very satisfactory results of fragmentation, there still
remains some underestimation of data for the yield of Li and Be
isotopes in $\calcium+\calcium$ system at 35 MeV/nucleon. A possible
solution of this problem will be the proper treatment of the
center-of-mass motions of clusters in a nucleus. Namely the diffusion
of the center-of-mass wave packet of a cluster, which often appears in
the fragmenting system, should be treated separately from the
diffusion of the wave packets of internal degrees of freedom of the
cluster. Such extensions will be made in the future work.

\acknowledgements
The numerical calculations were made by using the Vector Parallel
Processor, Fujitsu VPP500/28 of RIKEN.  We would like to thank Dr.\
K.\ Hagel for giving us the code of the experimental filter.

\end{document}